\documentclass[useAMS,usenatbib]{mn2e}

\usepackage{graphicx}
\usepackage{times}

\begin{document}
\newcommand{\heii}{He\textsc{ii}}
\newcommand{\hii}{H \textsc{ii}}
\newcommand{\hei}{He\textsc{i}}
\newcommand{\ion}[2]{[\textsc{#1}]#2}
\newcommand{\ha}{\ensuremath{\mathrm{H}\alpha}}
\newcommand{\hb}{\ensuremath{\mathrm{H}\beta}}
\newcommand{\nii}{{\ion{N ii}{6564}}}
\newcommand{\sii}{{\ion{S ii}{6716,6731}}}
\newcommand{\siii}{{\ion{S iii}{9069,9532}}}
\newcommand{\oiii}[1][5007]{{\ion{O iii}{#1}}}
\newcommand{\oii}{{\ion{O ii}{3727}}}
\newcommand{\ewha}{\ensuremath{\mathrm{EW(\ha)}}}
\newcommand{\dewha}{\ensuremath{\Delta \mathrm{EW}(\ha)}}
\newcommand{\ssfr}{SFR/M$_*$}
\newcommand{\HII}{H\textsc{ii}}
\newcommand{\apjs}{ApJS}
\newcommand{\apj}{ApJ}
\newcommand{\aj}{AJ}
\newcommand{\araa}{ARA\&A}
\newcommand{\apjl}{ApJL}
\newcommand{\nat}{Nature}
\newcommand{\pasp}{PASP}
\newcommand{\aap}{A\&A}
\newcommand{\mnras}{MNRAS}

\def\ltsima{$\; \buildrel < \over \sim \;$}
\def\simlt{\lower.5ex\hbox{\ltsima}}
\def\gtsima{$\; \buildrel > \over \sim \;$}
\def\simgt{\lower.5ex\hbox{\gtsima}}

%Main Document

\title[Spectral modelling of star-forming galaxies]{New insights into 
the stellar content and physical conditions of star-forming galaxies
at $z = 2 - 3$ from spectral modelling}

\author[J. Brinchmann, M. Pettini and S. Charlot]
       {Jarle Brinchmann,$^1$\thanks{Current address: Leiden Observatory, Leiden University, PO Box 9513, 2300 RA Leiden, the Netherlands}, Max~Pettini,$^2$ and St\'{e}phane Charlot$^3$\\
         $^1$Astrof\'{i}sica da Universidade do Porto, Rua das Estrelas, 4150-762 Porto, Portugal\\
         $^2$Institute of Astronomy, University of Cambridge, Madingley Road, Cambridge CB3 0HA, UK\\
         $^3$Institut d'Astrophysique de Paris,  UMR7095 CNRS, 
         Universit\'{e} Pierre \& Marie Curie, 
         98 bis Boulevard Arago, F-75014 Paris, France
      }

\date{Accepted 2008 January 2.  Received 2007 December 29; in original form 2007 July 13}
\pagerange{\pageref{firstpage}--\pageref{lastpage}}
\pubyear{2008}

\maketitle

\label{firstpage}

\begin{abstract}
We have used extensive libraries of model and empirical galaxy spectra
(assembled respectively from the population synthesis code of Bruzual
and Charlot and the fourth data release of the Sloan Digital Sky Survey)
to interpret some puzzling features seen in the spectra of
high redshift star-forming galaxies.
We show that a stellar He\,{\sc ii}\,$\lambda 1640$
emission line, produced in the expanding atmospheres of 
Of and Wolf-Rayet stars, should be detectable with an
equivalent width of $0.5-1.5$\,\AA\ in the integrated spectra
of star-forming galaxies, provided the metallicity is greater
than about half solar. 
Our models reproduce the strength of
the He\,{\sc ii}\,$\lambda 1640$ line measured in the
spectra of Lyman break galaxies for established values
of their metallicities.
With better empirical calibrations 
in local galaxies, this spectral feature has the potential of becoming 
a useful diagnostic of massive star winds at high, as well
as low, redshifts.

We also uncover a relationship in SDSS galaxies between 
their location in the [O\,{\sc iii}]/H$\beta$ vs. [N\,{\sc ii}]/H$\alpha$
diagnostic diagram (the BPT diagram) and their excess specific
star formation rate relative to galaxies of similar mass.
We infer that an elevated ionisation parameter $U$ is at the 
root of this effect, and propose that this is also the cause
of the offset of high redshift star-forming
galaxies in the BPT diagram compared to local ones.
We further speculate that higher electron densities and escape fractions
of hydrogen ionising photons may be the factors responsible 
for the systematically higher values of $U$ in the H\,{\sc ii}
regions of high redshift galaxies.
The impact of such differences on abundance determinations 
from strong nebular lines are considered 
and found to be relatively minor.
\end{abstract}


\section{Introduction}
\label{sec:introduction}

\begin{table*}
\begin{minipage}[c]{0.6\textwidth}
    \centering
    \caption{Adopted emission-line luminosities of individual Wolf-Rayet stars in
units of $10^{35}$ erg s$^{-1}$} 
    \begin{tabular}{@{}lccccc}
    \hline
    \hline
Line name & \multicolumn{2}{c}{WNE stars} & \multicolumn{2}{c}{WNL stars} & \multicolumn{1}{c}{OIf stars}\\
                  & $Z<0.2 Z_\odot$ & $Z \ge 0.2 Z_\odot$& $Z<0.2 Z_\odot$ & $Z \ge 0.2 Z_\odot$ \\
    \hline
He\,{\sc ii}\,$\lambda 4686$ &  1.7$^{a}$  & 8.4 &   4.3    &  24.7 & 1.3 \\
He\,{\sc ii}\,$\lambda 1640$ & 17    & 84  &   43     & 247 & 20 \\
N\,{\sc iii}\,$\lambda 4640$ &  0    & 0      & 4.0      & 6.3  & 0 \\
N\,{\sc v}\,$\lambda 4603-4620$   & 0.24$^{a}$ & 1.60     & 0 & 0 & 0 \\
    \hline
    \end{tabular}

\medskip
$^a$ Includes binaries, see Crowther \& Hadfield (2006).
    \label{tab:WR_line_lum}
\end{minipage}
\end{table*}

The spectra of galaxies, particularly at optical and ultraviolet (UV)
wavelengths, contain a wealth of information on their stellar
populations and interstellar media. The veritable explosion
in the size of galaxy surveys which we have witnessed in recent
years has been accompanied by the development of increasingly 
sophisticated models for the analysis and interpretation of 
their spectra (e.g. Kewley et al. 2001a; Bruzual \& Charlot 2003;
Heavens et al. 2004; Ocvirk et al. 2006).
The same analysis
tools are also increasingly being applied to the spectra
of galaxies at redshifts $z > 1$, even though
their detailed characteristics may well differ from
those of their counterparts at lower redshifts.
Considering, for example, star-forming galaxies,
it now seems well established that at high redshifts 
a larger fraction of the star formation activity took 
place in conditions similar to those encountered today 
in the relatively rare `luminous infrared galaxies' 
(e.g. Reddy et al. 2006 and references therein),
and that the transition from starburst dominated to a more
quiescent mode of star formation occurred at redshifts between
$z \simeq 2$ and 1 (e.g. Papovich et al. 2005).

It is thus reasonable to question to what extent 
diagnostic tools developed to interpret local galaxies
may need to be modified in order 
to decipher correctly the clues encoded in the spectra 
of high redshift galaxies. In this paper we use
state-of-the-art spectral synthesis models and 
observations of local galaxies to investigate two
apparent `anomalies' which have been noted in the 
spectra of high redshift star-forming galaxies.
The first is the common detection of the  
He\,{\sc ii}\,$\lambda 1640$ emission line with 
a width which is suggestive of an origin in the 
expanding atmospheres of luminous stars rather than 
in H\,{\sc ii} regions. This spectral feature is clearly seen
in the composite rest-frame UV spectrum of more than 
800 Lyman break galaxies at $z \simeq 3$ constructed
by Shapley et al. (2003). As those authors point out,
its strength is underestimated by continuous
star formation models generated by the Starburst99 code
(Leitherer et al. 1999), 
particularly at sub-solar metallicities. 
Locally, the He\,{\sc ii}\,$\lambda 1640$
emission line is prominent in the UV spectra 
of starburst galaxies, but only during a brief period when 
the number of Wolf-Rayet (W-R) stars is at a maximum 
(e.g. Chandar, Leitherer \& Tremonti 2004),
so that its strength in the composite spectrum of hundreds of
galaxies which span a wide range of ages, from $10^7$ to $10^9$ years
(Shapley et al. 2001; Erb et al. 2006b), is at first sight puzzling. 
Here we reassess this `problem' using the most 
recent models and observations of Wolf-Rayet stars in 
galaxies of differing metallicities, and taking into
account a range of star formation histories.

The second issue concerns the indication that star-forming
galaxies at $z \simgt 1$ may exhibit systematically higher 
ratios of collisionally excited to recombination lines 
than  present-day  H\,{\sc ii} regions
(Shapley et al. 2005; Erb et al 2006a; Kriek et al. 2007). 
Although the data samples are still small, most of the 
galaxies at $z \simgt 1$ where the relevant
emission lines have been measured, are offset---relative 
to lower $z$ galaxies---in the 
[O\,{\sc iii}]$\lambda 5007$/H$\beta$ vs. 
[N\,{\sc ii}]/H$\alpha$ diagram
first proposed by Baldwin, Phillips 
\& Terlevich (1981, BPT) as a way of
classifying emission line regions and their sources
of ionisation/excitation. 
A variety of different physical processes may be
at the root of such differences, as discussed
by Shapley et al. (2005, see also Liu et al 2008); we investigate this question
further here with detailed modelling and comparison
with the large body of data on local galaxies provided
by the Sloan Digital Sky Survey (SDSS).

Where relevant, we have adopted today's consensus
cosmology with density parameters
$\Omega_{\rm M}=0.3$, $\Omega_\Lambda=0.7$ and 
a Hubble parameter $H_0=70$\,km~s$^{-1}$~Mpc$^{-1}$. 
We have also adopted throughout the stellar initial mass function
(IMF) of Chabrier (2003), unless otherwise specified.

\section{Origin of the He\,{\sc ii}~$\lambda 1640$ emission line in 
the integrated spectra of high redshift
star-forming galaxies}

\subsection{Theoretical Models}
\label{sec:theor_models}

%We use as a starting point a library of galaxy
%model spectra produced by $\sim 10^5$ stochastic realisations 
%of the population synthesis code by 
%Bruzual \& Charlot (2003, BC03). 
%The details of this grid of models have been described by
%Gallazzi et al. (2005) and Salim et al. (2005);
%we refer the interested reader to those
%papers for a full discussion of how they are generated. 
%Briefly, the star formation history is characterised
%by two components: an underlying continuous star formation
%with an exponentially declining rate 
%determined by the parameter $\tau$
%(${\rm SFR} \propto \exp [t_{\rm sf}/\tau]$),
%on which random bursts can be 
%superimposed.
%Dust obscuration is included so that the $V$-band
%absorption optical depth $\tau_V$ across 
%the full model grid is consistent with the properties of the 
%SDSS galaxies, although for the present purposes 
%the dust content is not very important.

We use as a starting point the spectral evolution 
models by Bruzual \& Charlot (2003, BC03). 
We have adopted the Padova 1994 tracks (see BC03 for details) 
for the stellar evolution but use Geneva 'high-mass-loss' tracks 
(Meynet et al. 1994) for the Wolf-Rayet phase. 
These provide a good match to the observational data 
and agree with tracks for stellar evolution which take
into account the effects of stellar rotation 
(Meynet \& Maeder 2005; Vazquez et al. 2007).
For the work described in this paper, we have made one significant
addition to the Bruzual \& Charlot (2003) models by including the
predictions of the strengths of UV and optical emission lines due to
W-R stars from the stellar evolution models by Schaerer \& Vacca
(1998).  It is important to emphasise here that the absolute
calibration of the luminosity of the He\,{\sc ii}\,$\lambda 1640$ line
(of particular interest to our present investigation) is based on a
still rather scant body of measurements. Schaerer \& Vacca (1998) used
reference luminosities for the optical He\,{\sc ii}\,$\lambda 4686$
emission line from the best empirical compilations available a decade
ago and then adopted appropriate scaling factors to predict the
luminosity of the $\lambda 1640$ line (for WN stars; for WC and WO
stars the reference line is C\,{\sc iv}\,$\lambda 5808$).  More
recently, Crowther \& Hadfield (2006) have reexamined these
calibrations with a more comprehensive sample of W-R spectra and
better corrections for reddening, and concluded that: (a) the He\,{\sc
  ii}\,$\lambda 4686$ line has a higher luminosity in metal-poor WNE
stars than the value adopted by Schaerer \& Vacca (1998); (b) the
ratio of the luminosities in the two He\,{\sc ii} lines, $L(\lambda
1640)/L(\lambda 4686)$, is 20\% larger than assumed by Schaerer \&
Vacca; and (c) there is a metallicity dependence, not only in the
ratio of W-R to O-type stars, but also in the sense that W-R stars
with lower metallicity have lower luminosity He\,{\sc ii} lines.

We have modified the Schaerer \& Vacca (1998) models as implemented in
Starburst99 to reflect these changes in the overall calibrations of
the He\,{\sc ii} emission line luminosities.  Specifically, for WN
stars we have adopted the line luminosities listed in
Table~\ref{tab:WR_line_lum}, while for WC stars we have retained the
calibrations by Schaerer \& Vacca (1998).  Following Crowther \&
Hadfield (2006), we have made a distinction between stars with
metallicities lower and higher than 1/5 solar (this being
approximately the metallicity of the Small Magellanic Cloud).
As discussed below we have also adopted 
the He\,{\sc ii}\,$\lambda 4686$ luminosity for WN class 5--6 
as representative of the WNL luminosity. 

For OIf\footnote{Of stars are O-type stars where
He\,{\sc ii}\,$\lambda 4686$ is found in emission
due to their relatively stronger winds compared
to those of main-sequence O stars, where the line
is in absorption.} stars we have adopted the 
luminosities given in Table~\ref{tab:WR_line_lum},  
kindly provided by P. Crowther (2007, priv.\  comm). 
The inclusion of these stars does make a difference at very
early stages in a starburst and particularly for low metallicities
where the WR phase is much less prominent. 
At $Z=0.2Z_\odot$ the inclusion
of OIf stars increases the 
He\,{\sc ii}\,$\lambda 1640$ luminosity by
$\sim 50$\% and larger gains are seen at 
lower metallicities.

We have also adopted a ratio of 10 between the luminosity of 
He\,{\sc ii}\,$\lambda 1640$ and He\,{\sc ii}\,$\lambda 4686$ for 
both WNL and WNE stars.  
We then used Starburst99 to calculate single-stellar
populations from these models, using the same IMF and stellar tracks
as used by Bruzual \& Charlot (2003).  These single-stellar
populations SSP were finally interpolated onto the same time grid as
used by Bruzual \& Charlot (2003) to allow the calculation of
arbitrary star formation histories.

Even with these revisions, however, substantial uncertainties remain
in the modelling. They have two separate causes: the association of
observational data with theoretical models and the distribution of
observed fluxes. The connection between observational data and
theoretical tracks is non-trivial essentially because the observed
features originate in the stellar wind and the models focus on the
stellar interior. Thus spectroscopic classifications of WR stars do
not necessarily map naturally onto the evolutionary model
parameters --- see for instance the discussion by Foellmi et al. (2003)
and Meynet \& Maeder (2003). 
Furthermore, the currently available libraries 
of stellar tracks (Meynet \& Maeder 2003) do not include rotation 
(see Vazquez et al. (2003) for a partial implementation) 
and the population evolution modes do not include a 
binary channel (see however Han et al. 2007) which may be of crucial 
importance for Wolf-Rayet formation at low metallicity 
(Foellmi et al. 2003; Crowther 2007).

The second problem arises because there seems to be a wide spread in
the luminosities of the He\,{\sc ii} lines measured in W-R stars of
even the same sub-type, and the available data are still too few to
ascertain the form of the distribution function of, for example,
$L(\lambda 4686)$ for a given W-R spectral type or sub-type (see
Figure 2 of Crowther \& Hadfield 2006). The luminosities
span such a large range that any predictions are sensitive to the
extreme, and hence rare, values. Thus, measurements in large
samples of W-R stars---not yet assembled---are required to 
characterise the distribution adequately.
This is indeed the reason why we have adopted the WN class 5--6 as
representative of the WNL luminosity, since this is the best-sampled
WNL class.

In summary, until more comprehensive observational libraries and further
theoretical work on the line fluxes in W-R stellar atmospheres become
available, we have to accept the fact that even the best evolutionary
synthesis models probably cannot predict the luminosities of W-R
emission lines with an accuracy better than a factor of $\sim 2$--$3$
(based on the empirical spread of the luminosities of these lines and
differences in mapping between empirical data and evolutionary models).

\begin{figure}
  \vspace*{-0.05cm}
  \centering
  {\hspace*{-0.8cm} 
    \includegraphics[angle=90,width=93mm]{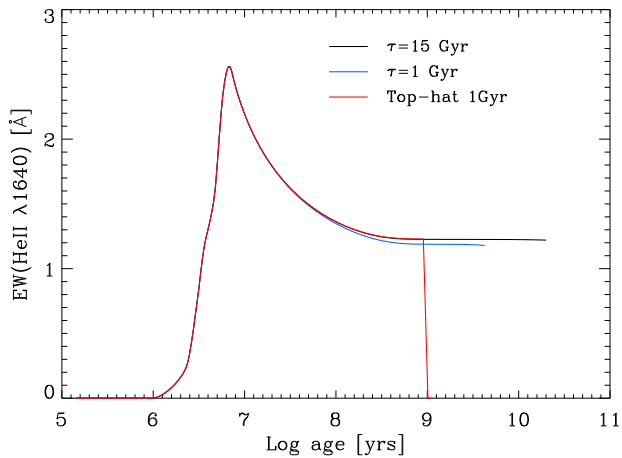}
  }
  \caption{Time evolution of the equivalent width of the 
   He\,{\sc ii}\,$\lambda 1640$ emission line for three
    different star formation histories. The black line is for a 
    continuous star formation model in which the star formation rate
    declines slowly, with an $e$-folding time $\tau=15$\,Gyr.
    The blue line is for a faster decline ($\tau=1$\,Gyr), while the
    red line is for a single burst of star formation at a fixed rate
    and lasting 1\,Gyr, after which the star formation abruptly ceases.
    } 
  \label{fig:ew1640_vs_sfh}
\end{figure}

\subsection{Synthesising the He\,{\sc ii}\,$\lambda 1640$ emission line:
simple star formation histories}
\label{sec:model_HeII}

With these limitations in mind, we now consider the 
strengths of the He\,{\sc ii}\,$\lambda 1640$ lines 
generated by our grid of models.
As a general consideration we recall that only 
the hottest stars (or a power-law continuum 
as produced for example by 
an active galactic nucleus) can photoionize 
He$^+$ to create nebular He\,{\sc ii} emission 
since this requires photons with energies in excess of 
54.4\,eV. However both early O supergiants and 
Wolf-Rayet stars have sufficient numbers of photons 
with energy above 24.6\,eV to create \emph{stellar} He\,{\sc ii} emission. 

Given the short lifetimes of these stars, the stellar 
He\,{\sc ii}\,$\lambda 1640$ line would be 
present for only a few Myr following a burst 
of star formation. However in the more realistic case of a 
protracted star formation episode, this narrow `step-function' 
becomes smoothed out, as we now explore.

\begin{figure}
  \vspace*{-0.05cm}
  \centering
%  {\hspace*{-0.8cm} 
%  \includegraphics[angle=90,width=93mm]{ew1640_evolution_const_vs_z.ps}}
  {\hspace*{-0.8cm} 
    \includegraphics[angle=90,width=93mm]{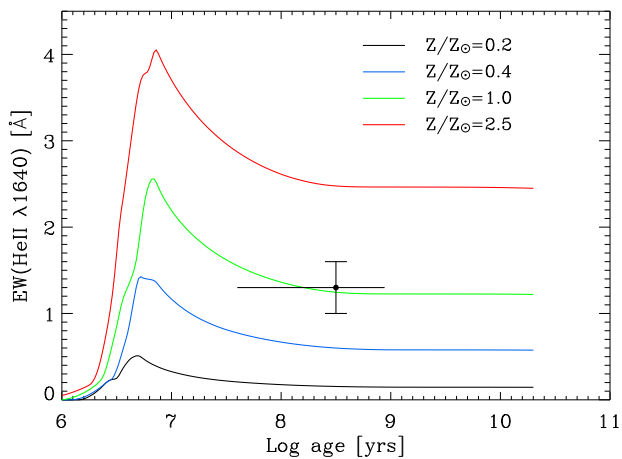}}
    \caption{Time evolution of the equivalent width of the 
    He\,{\sc ii}\,$\lambda 1640$ emission line for four difference
    metallicities, as indicated. All four curves refer to the 
    model with slowly declining star formation rate ($\tau=15$\,Gyr). 
    The dot at $\log t=8.5$ 
    is the value of EW(He\,{\sc ii}\,$\lambda 1640$) measured
    from the composite spectrum of 811 Lyman break 
    galaxies published by Shapley et al. (2003), while
    the error bar shows the inter-quartile range of their ages (Shapley et al. 2001). 
    The measured value of EW(He\,{\sc ii}\,$\lambda 1640$) is
    reproduced by our models with metallicities $\mathbf{Z = 0.75 - 1.5\,Z_{\odot}}$,
    somewhat higher than, but still consistent with, current estimates of the metallicity 
    of LBGs. } 
  \label{fig:ew1640_vs_Z}
\end{figure}

Figure~\ref{fig:ew1640_vs_sfh} illustrates the time evolution 
of the equivalent width of the He\,{\sc ii}\,$\lambda 1640$
emission line for three different star formation histories.
The black line shows the case of a nearly constant
star formation rate (approximated by an exponentially
declining rate of star formation with a very long
decay time of 15\,Gyr); the blue line 
is for a more rapidly declining star formation rate
($\tau=1$\,Gyr); and the red line is for a discrete 
burst of star formation at a constant rate lasting
1\,Gyr (after which the star formation is abruptly switched off).  
For the moment we keep the metallicity constant at the
solar value. 

\begin{figure*}
  \centering
%  {\hspace{-0.35cm}\includegraphics[angle=90,width=84mm]{ew1640_burst_on_exp.ps}}
  {\hspace{-0.25cm}\includegraphics[width=174mm]{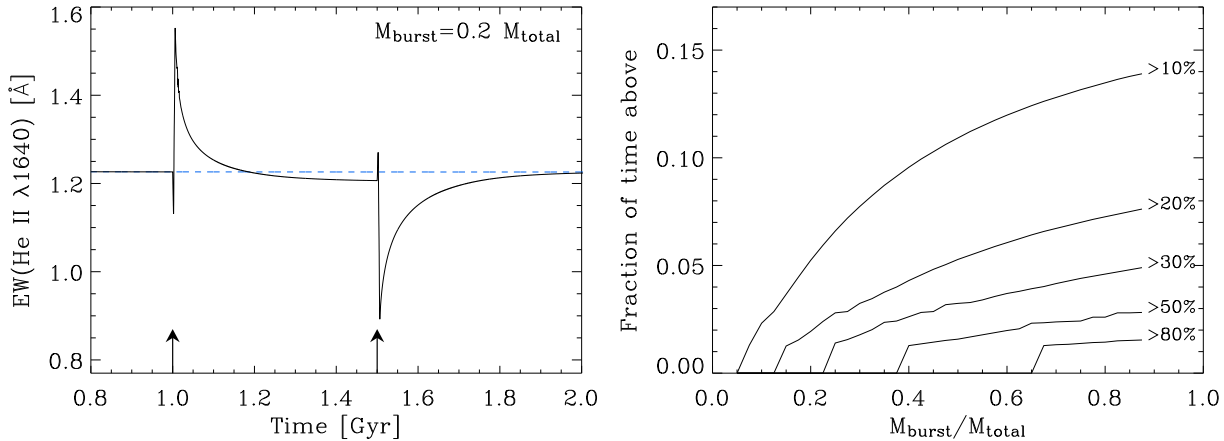}}
  \caption{\textit{Left panel:}~An illustration of the effect of a burst of star formation
   on the equivalent width of the He\,{\sc ii}\,$\lambda 1640$ emission
   line. The burst is superimposed on an underlying level of
   near-constant star formation which gives 
   EW($\lambda 1640) \simeq 1.2$\,\AA\ (shown by the horizontal
   short-dash line). At $t_{\rm sf}=10^9$ years, a burst of duration
   5$\times 10^8$ years, building up 20\% of the stellar mass, 
   is superposed (the arrows indicate the start and 
   end of burst).  
   \textit{Right panel:} The effect on EW($\lambda 1640$) 
   of a burst of star formation superposed on a pre-existing 
   stellar population forming stars at a constant rate.  
   The $x$-axis shows the strength of the burst, 
   as measured by the fraction of stellar mass contributed 
   by the burst to the final total stellar mass. 
   The $y$-axis shows the fraction of time that the 
   EW($\lambda 1640$) is increased by a given 
   percentage (indicated at the right of each curve) 
   above the value for constant star formation.} 
  \label{fig:heii_burst}
\end{figure*}

%\begin{figure}
%  \centering
%%  {\hspace{-0.25cm}\includegraphics[angle=90,width=84mm]{fraction_ew1640_above.ps}}
%  {\hspace{-0.25cm}\includegraphics[angle=90,width=84mm]{fig4.ps}}
%  \caption{The effect on EW($\lambda 1640$) of a burst of star 
% formation superposed on a pre-existing stellar population forming stars at a constant rate  
% The $x$-axis shows the strength of the burst, as measured by the fraction 
% of stellar mass contributed by the burst to the final total stellar mass. 
% The y-axis shows the fraction of time that the EW($\lambda 1640$) 
% is increased by a given percentage (indicated at the right of each curve) 
% above the value for constant star formation.} 
%  \label{fig:fraction_time_burst}
%\end{figure}

These three examples were chosen to illustrate
star formation histories which may be appropriate
to composite spectra of \emph{many} galaxies
(such as the LBG average constructed by Shapley et al. 2003)
where the individual characteristics of the component
spectra are smoothed out. These models are therefore 
not applicable to individual galaxies, caught at a
particular stage in their evolution. 
It is worth pointing out here that during short
bursts of star formation, EW(He\,{\sc ii}\,$\lambda 1640$)
can reach values two to three times higher than those
shown in Figure~\ref{fig:ew1640_vs_sfh}; such models
would be able to reproduce the observations
of He\,{\sc ii}\,$\lambda 1640$ in individual
star-forming galaxies at low
(e.g.\ Gonzalez-Delagado et al. 1999; Leitherer et al. 2002; Chandar et al. 2004),
as well as high (e.g.  Lowenthal et al. 1997; Kobulnicky \& Koo 2000),
redshifts.

Returning to Figure~\ref{fig:ew1640_vs_sfh},
it can be seen that in all three cases shown, 
the equivalent width of the He\,{\sc ii} emission
line reaches a peak value only a few million years after the onset of
star formation, when the number of W-R stars is at a maximum. While
this behaviour was expected, what has perhaps not been fully
appreciated until now is that, following this W-R bright phase, the
quantity EW($\lambda 1640$) settles to a plateau value of $\simeq
1.5$\,\AA\ which is maintained until the star formation rate
essentially stops.  The physical reason for this plateau is that the
underlying UV continuum at 1640\,\AA\ is also due to massive stars
with short evolutionary timescales (albeit not as massive, nor as
short-lived as the W-R stars, the luminosity-weighted
$T_{\mathrm{eff}}$ at $\lambda=1640$\,\AA\ is 25,000\,K)  rather than
being set by the previous, longer-term, star formation history. Thus
the ratio between the luminosity in the He\,{\sc ii} emission line and
in the continuum---which the EW measures---tends to a constant value
after about $10^8$ years, when the number of OB stars has stabilised.  

Next we consider the effect of metallicity on the predicted value of
EW($\lambda 1640$), limiting ourselves to the slowly declining star
formation rate case with $\tau=15$\,Gyr.  As can be readily
appreciated from Figure~\ref{fig:ew1640_vs_Z}, while the general form
of the time evolution of EW($\lambda 1640$) is common to all four
values of metallicity considered (from $Z = 2.5\,Z_{\odot}$ to $Z =
1/5\,Z_{\odot}$), the peak and plateau values of EW($\lambda 1640$)
depend sensitively on metallicity, in the sense that the He\,{\sc ii}
emission line is considerably weaker at subsolar metallicities (and
stronger when $Z > Z_{\odot}$).  Such a marked dependence is due to
two effects. First, and most important, the number of massive stars
which evolve to the W-R stage is substantially reduced at sub-solar
metallicities, as evidenced by the reduced ratio of W-R to O-type
stars---as well as the change in the W-R sub-type
distribution---between the Milky Way and the Magellanic Clouds (e.g.
Crowther 2007 and references therein).  Second, as mentioned above,
there is evidence that in individual W-R stars of a given sub-type the
luminosity of the He\,{\sc ii} lines is lower at lower metallicities
(Crowther \& Hadfield 2006).  The metallicity dependence of the winds
from the W-R progenitors is presumably at the root of these
differences (Vink \& de Koter 2005).

Also shown in Figure~\ref{fig:ew1640_vs_Z} is the value   
EW($\lambda 1640) = 1.3 \pm 0.3$\,\AA\ 
measured from the composite of spectrum of 811 Lyman
break galaxies published by Shapley et al. 
(2003).\footnote{The spectrum is available in digital form at 
http://www.astro.princeton.edu/$\sim$aes/}
In order to measure the equivalent width, we normalised
the spectrum according to the prescription by
Rix et al. (2004); the error quoted reflects the uncertainties
in both the continuum placement and the width of the emission
line.  From our models, this value of 
EW($\lambda 1640$) is that expected for star-forming
galaxies where star formation has been progressing for
longer than $\sim 100$\,Myr and the metallicity is roughly
between 0.75 and 1.5 times solar. 
For comparison, we estimate that the typical
metallicity of the Lyman break galaxies 
in the composite of  Shapley et al. (2003)
is in the range $Z_{\rm LBG} \simeq 0.3 - 1 Z_{\odot}$,
based on the determinations by 
Pettini et al. (2001) and Erb et al. (2006a; see also Appendix A),
while their median age is $\sim 320$\,Myr (Shapley et al. 2001).
Thus, while the models seem to underestimate somewhat the
equivalent width of the He\,{\sc ii}\,$\lambda 1640$ emission line,
it is nevertheless encouraging that they come so close,
given the current uncertainties in 
the absolute calibrations of the luminosities of the He\,{\sc ii} lines,
as emphasised above.
We also point out that our models match simultaneously
the equivalent width of the C\,\textsc{iv}\,$\lambda 1550$ P-Cygni
line, unlike previous attempts to account for the 
He\,{\sc ii}\,$\lambda 1640$ emission by appealing to 
a top-heavy IMF (e.g.\  Shapley et al. 2003;  Chandar et al. 2004).
Furthermore, in our models the luminosity-weighted width of the 
optical He\,{\sc ii}\,$\,\lambda 4686$ line of $\sim 1500$--$2000$\,km~s$^{-1}$. 
A similar width is expected for He\,{\sc ii}\,$\lambda 1640$,
in good agreement with the FWHM($\lambda 1640$)\,$\sim 1500$\,km~s$^{-1}$
of this spectral feature in the Shapley et al. (2003) composite spectrum.
This gives added confidence to the identification of this line as originating in the winds
of W-R and OIf stars.

\subsection{The effect of bursts of star formation}
\label{sec:bursts_effect}

We now briefly consider the effect on the equivalent 
width of the He\,{\sc ii}\,$\lambda 1640$ line of a less
simplistic model for the star formation activity,
by adding a burst of star formation to an underlying
continuous mode. We illustrate this composite scenario
in the left panel of Figure~\ref{fig:heii_burst}, where the burst is 
introduced as a step function at time 
$t_{\rm sf} = 1$\,Gyr  and lasts for 0.5\,Gyr. 
The underlying continuous mode is, as before,
a slowly declining one with $\tau = 15$\,Gyr,
and the strength of the burst is such that 
it accounts for 20\% of the stellar mass of the system 
at the end of the burst.

Referring to the figure, we see that, following
the burst, the equivalent width EW($\lambda 1640$)
increases by about 25\% due to the enhanced contribution
from the Of and W-R stars created in the 
burst\footnote{This increase is preceded by a very short-lived
period when EW($\lambda 1640$) actually decreases by a few percent.
We are unsure as to whether this effect is real, but it may be
due to the relatively low contribution to $L(\lambda 1640)$ by
the stars with the very highest masses in our model. In any case,
this minor feature is unimportant for our purposes here.}.
It subsequently declines to the pre-burst value on a timescale
of $\sim 200$\,Myr as the contribution to the continuum from
lower-mass O and B-type stars without a He$^+$ zone grows.
After the end of the burst, at $t_{\rm sf} = 1.5$\,Gyr, 
the EW drops until all the massive stars formed in the
burst have faded away, and then rises back to
the pre-burst equilibrium value. 
The decline in EW after the start of the
burst and its increase after the end of the burst have
the same shape because the same stellar
populations contribute (with the same timescales) 
in both cases.

The features seen in  Figure~\ref{fig:heii_burst}
are generic to this kind of composite scenario,
although the details would change with the exact parameters
of the burst (and the sharp transitions in Figure~\ref{fig:heii_burst}
would of course be smoothed out if one considered a series of bursts
or smoother onset/termination of star formation).
The important point, however, is that, while bursts
of star formation do have an effect on the equivalent
width of the He\,{\sc ii}\,$\lambda 1640$ line,
the effect is relatively modest and short-lived.
We quantify this conclusion in the right-hand panel of 
Figure~\ref{fig:heii_burst} which shows
the fraction of time during which the quantity
EW($\lambda 1640$) is increased by a given percentage
over the equilibrium value following the onset of a
500\,Myr long burst of star formation. The effect 
is at the $20 - 30$\% level for less than 10\% of
time (that is, for less than  50\,Myr); higher perturbations are even more
short-lived.
Such variations are small compared
with the more important uncertainty in the
overall calibration of the luminosity of the 
He\,{\sc ii} lines resulting from the poor 
empirical sampling of the spectra of W-R stars
in nearby galaxies. 
Of course, the increase in EW($\lambda 1640$)
following a burst can be made larger
by increasing the 
contrast of the burst over the underlying continuous
star formation activity (see right-hand panel of Figure~\ref{fig:heii_burst}).
Thus, if a strong burst 
takes place after a period of low star formation activity
(lasting longer than the lifetimes of the OB stars
contributing to the UV continuum), then the increase
in EW($\lambda 1640$) will be much larger than the 
$\sim 25$\% increase appropriate to the model shown in
Figure~3. Nevertheless, even in these circumstances,
the period during which EW($\lambda 1640$) is abnormally
high is brief, and the equivalent width of the line settles
to the equilibrium value on a $\sim 10^8$\,year timescale.

Summarising the conclusions of this section, our models
show that the presence of a detectable 
He\,{\sc ii}\,$\lambda 1640$ emission line 
in the spectra of galaxies actively forming stars
need not be surprising and requires no exotic stellar populations (cf.\ Jimenez \& Haiman 2006). Provided the metallicity is
higher than about half solar, we expect that
this emission line should have an equivalent 
width EW($\lambda 1640) \sim 0.5$\,\AA\ or greater
while star formation is proceeding. 
The equivalent width
settles to an equilibrium value because its
value is determined from stars at the upper end
of the initial mass function; perturbations from
this equilibrium value are short-lived because 
they are determined by a temporary excess, or deficit,
of Of and W-R stars compared to O and early B stars.
For this same reason, EW($\lambda 1640$) responds
sensitively to changes in metallicity, since it
is the metallicity that apparently determines the
ratio of W-R to O stars and, to a lesser extent,
the luminosity of the He\,{\sc ii}\,$\lambda 1640$ line
in a given W-R star. Considering the paucity of current
data on the strength of this emission line in stars
of different W-R sub-types and metallicities,
it is perhaps fortuitous that our
models reproduce within a factor of $\sim 2$ the value of 
EW($\lambda 1640$) in the composite spectrum of
Lyman break galaxies at $z \simeq 3$,
for current estimates of their metallicities. 
Nevertheless, the agreement is encouraging.
There are good prospects that, 
as more data are gathered on the luminosity 
$L(\lambda 1640)$ in local galaxies with different
metallicities and star formation histories, 
we may be able to use the He\,{\sc ii} line to
investigate the properties of stellar winds
in galaxies at high redshifts.

\begin{figure}
  \centering
%  {\hspace*{-0.9cm}\includegraphics[angle=90,width=104mm]{bpt_diagram_with_fit.ps}}
  {\hspace*{-0.9cm}\includegraphics[angle=90,width=104mm]{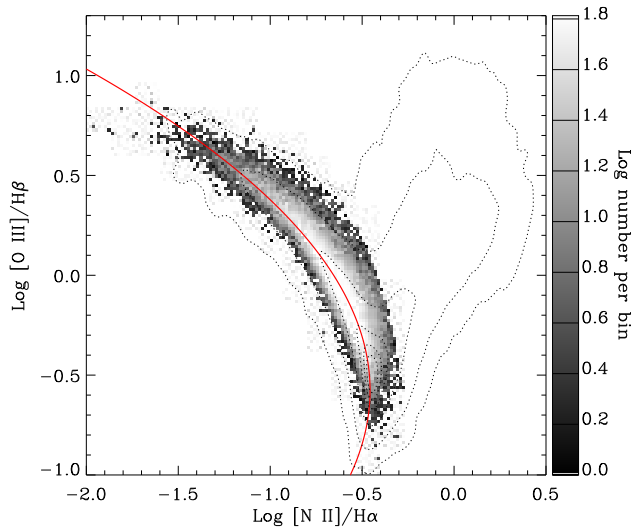}}
  \caption{The BPT diagram of actively star-forming galaxies from 
   the SDSS DR4. The number of galaxies in each bin of
   $\log$\,[O\,{\sc iii}]/H$\beta$ vs. $\log$\,[N\,{\sc ii}]/H$\alpha$ 
   is shown on a logarithmic grey scale
   indicated on the right of the diagram. 
   The analytic fit to
   the red ridge line is given by eq.\,(1) (see text).
   The dotted contours enclose, respectively, 5, 10, 20, 50, 90 and 99\%
   of all galaxies in DR4 (in all three 
   classes: Star-Forming, AGN, and Composite) 
   in which the four emission lines are detected at a 
   greater than $5 \sigma$ significance
   level.
   }
  \label{fig:bpt_dr4}
\end{figure}

\begin{figure}
  \vspace*{-0.1cm}
  \centering
%
%{\hspace*{-1.25cm}\includegraphics[angle=90,width=108mm]{bpt_with_highz.ps}}
  {\hspace*{-1.25cm}\includegraphics[angle=90,width=108mm]{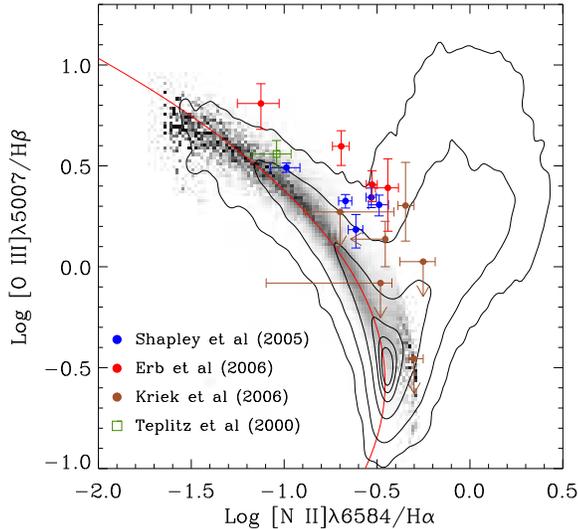}}

  \caption{SDSS and high redshift galaxies in the [O\,{\sc iii}]/H$\beta$
    vs. [N\,{\sc ii}]/H$\alpha$ diagnostic diagram. Galaxies at
    $z > 1$ are from the sources indicated in the lower left-hand corner
    of the diagram. The grey scale used here to represent the SDSS galaxies
    was constructed by normalising the distribution function in bins along
    the $x$-axis, so that the total number of galaxies in each bin is the 
    same. This representation has the advantage of bringing out more clearly
    the structure of the locus occupied by star-forming galaxies in the SDSS.
    Contours are as in Figure~\ref{fig:bpt_dr4}.
    } 
  \label{fig:bpt_with_highz}
\end{figure}

\section{The [O\,{\sc iii}]\,$\lambda 5007$/H$\beta$ 
vs. [N\,{\sc ii}]$\lambda 6583$/H$\alpha$
diagram at low and high redshift}
\label{sec:bpt_hiz_loz}

Since the seminal paper by Baldwin et al. (1981), this 
diagnostic diagram has been used extensively to separate
star forming galaxies from LINERS, Seyfert 2 galaxies, 
and mixed AGN/stellar ionisation systems 
(e.g. Veilleux \& Osterbrock 1987;
and, more recently, Kauffmann et al. 2003; 
Brinchmann et al. 2004 among many others).

In order to compare galaxies at high redshift with local
ones, we use the fourth data release (DR4) of the SDSS
(Adelman-McCarthy  et al. 2006; York et al. 2000). 
SDSS spectra were obtained using a fibre spectrograph 
with 3\,arcsec wide fibre apertures and 
we re-analyse the galaxy spectra with the pipeline reduction 
described by Tremonti et al. (2004). In this sample
stellar masses were derived using the precepts of
Kauffmann et al. (2003), and strong-line oxygen abundances
and star formation rates determined with the Bayesian method
discussed by Brinchmann et al. (2004).
The overall properties of this sample are similar to
those of the smaller SDSS DR2 sample of
Brinchmann et al. (2004), whose criteria we have adopted 
to classify the galaxy emission line spectra as 
dominated by star formation (the SF class), AGN, or
a mix of the two (the Composite class). 
Here we limit ourselves to galaxies in the SF class,
which are plotted on the BPT diagram in Figure~\ref{fig:bpt_dr4};
we have included all galaxies in which the four emission
lines in question are detected with a signal-to-noise ratio
$S/N > 5$; there are 85748 galaxies in DR4 satisfying these
criteria. 
We elected not to consider the Composite class 
because a secure identification of the main ionisation mechanism can only 
be made for a small subset of objects in that class.
However, Liu et al. (2008) have examined these more extreme 
galaxies in a recent analysis which complements that 
presented here.

A notable feature of the BPT diagram is 
that star-forming galaxies
fall in a narrow region of the 
[O\,{\sc iii}]/H$\beta$ vs. [N\,{\sc ii}]/H$\alpha$ plane,
around a ridge line well represented by the quadratic equation:
%
% Created with:  fit_bpt_ridgeline, sc, lgm, class_zw, res=res
%
% For DR6: -0.591 y^2 - 0.674 y -0.658 [it is indistinguishable]
% 
\begin{equation}
  \label{eq:bpt_ridge}
  x = -0.596 y^2 - 0.687 y - 0.655
\end{equation}
where $x = \log$\,[N\,{\sc ii}]/H$\alpha$ and 
$y = \log$\,[O\,{\sc iii}]/H$\beta$.
In determining this best-fit line we have excluded
galaxies with stellar mass surface density
$\mu_{\star} > 3  \times 10^8\,M_{\odot}$~kpc$^{-2}$.
Such high stellar mass surface densities 
are characteristics of luminous bulges;
as shown by Kauffmann et al. (2003), their
removal from the sample is effective in 
excluding galaxies with low levels of AGN activity.

As mentioned in the Introduction, star-forming 
galaxies at $z \simgt 1$ seem to be offset 
significantly from this ridge line. Although 
all four emission lines have been measured 
in only a handful of high redshift galaxies so far
(due to the limitations of current near-infrared
spectrographs), the offset from the large body of SDSS
measurements is readily apparent in 
Figure~\ref{fig:bpt_with_highz}. 
This offset can be quantified with two parameters:
the distance \textit{along} the ridge line defined 
by eq.\,(1), $d_{\rm A}$ (defining arbitrarily the
zero point to be at $x_0 = -0.61$, $y_0 = -1.1$), and
the distance perpendicular to the ridge line, $d_{\rm P}$.
The high redshift galaxies included in 
Figure~\ref{fig:bpt_with_highz} are UV-luminous galaxies
at $z \simeq 2.2$ from Erb et al. (2006a), selected
according to the `BX' criteria of Steidel et al.\ (2004), and  including
unpublished data for Q2343-BX474 (Erb 2007, priv.\ communication);
the star-forming galaxies at $z = 2.0 - 2.6$ from
the $K$-band selected sample studied by Kriek et al. (2007),
excluding those classified as AGN; the star-forming
galaxies at $z = 1.0 - 1.5$ from the DEEP2 galaxy redshift
survey observed by Shapley et al. (2005); and the gravitationally
lensed $z = 2.7276$ Lyman 
break galaxy MS\,1512--cB58 (Teplitz et al. 2000). 
This still meagre sample
will undoubtedly increase significantly in the near future (e.g.\ Liu et al 2008),
as the multi-object and multi-band capabilities of near-IR 
spectrographs on large telescopes improve. The purpose
of our investigation here is to identify the physical
processes which are responsible for the offset of high-$z$
galaxies in the BPT diagram.

\subsection{Star formation history and the BPT diagram}
\label{sec:sfr_variation}

\begin{figure}
  \vspace*{-0.15cm}
  \centering
%  {\hspace*{-0.70cm}
%  \includegraphics[angle=90,width=98mm]{bpt_delta_log_ewha_mstar_n2s2_bins.ps}} 
  {\hspace*{-0.70cm}
  \includegraphics[angle=90,width=98mm]{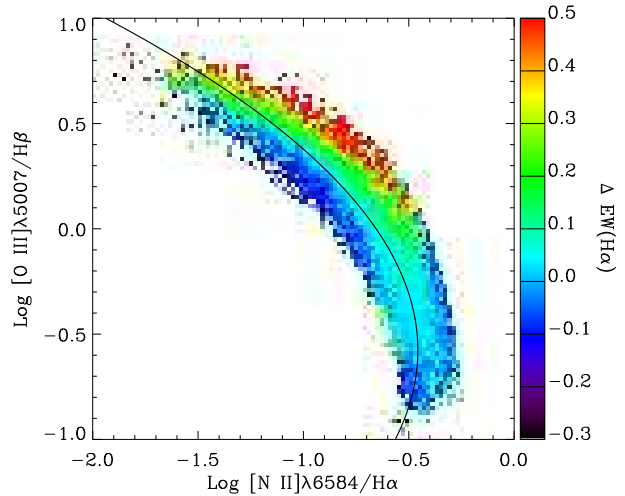}} 
  \caption{Variation of $\Delta{\rm EW(H}\alpha)$ in the BPT plane. 
  The parameter $\Delta{\rm EW(H}\alpha)$, which measures 
  the excess or deficit H$\alpha$
  equivalent width compared to galaxies of similar mass, 
  is clearly related to the distance of a galaxy from the  
  ridge line (shown here in black).
  } 
  \label{fig:bpt_variation}
\end{figure}

\begin{figure*}
  \centering
%  {\hspace{-0.25cm}\includegraphics[width=174mm]{ewha_bpt_comparison_hiz.ps}}
  {\hspace{-0.25cm}\includegraphics[width=174mm]{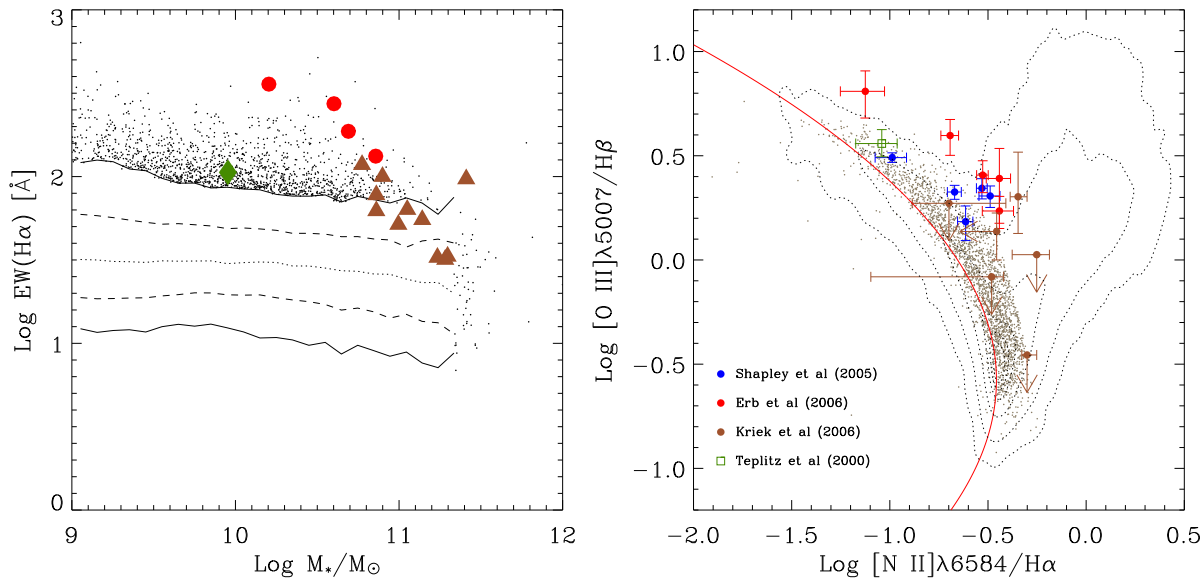}}
  \caption{\textit{Left panel:}~The equivalent width of the H$\alpha$ emission line
   as a function of stellar mass. The dotted line shows the 
   median trend for SDSS DR4 star forming galaxies. The dashed
   and continuous lines enclose respectively 68\% and 95\% of
   the galaxies. The lines have been drawn by binning galaxies
   in intervals of stellar mass chosen so that each bin includes 
   2500 galaxies. Galaxies at the extremes of the distribution, 
   that is with values of EW(H$\alpha$) above the 97.5\% limit 
   and with stellar masses $M_{\star} > 2 \times 10^{11}\,M_{\odot}$
   are plotted individually with small black dots. Coloured symbols
   refer to high redshift galaxies, 
   as indicated in the legend in the right hand figure. 
   The solid triangle at $\log M_{\star}/M_{\odot} \simeq 10$ 
   shows the location of the galaxy MS\,1512-cB58.
   \textit{Right panel:}~Location on the BPT diagram of:
   (a) the SDSS
   galaxies plotted individually with small dots in the left panel,
   and (b) the high redshift galaxies. Red ridge line and contours
   as in Figures~\ref{fig:bpt_dr4} and \ref{fig:bpt_with_highz}.
   }
  \label{fig:bpt_hiz_explanation1}
\end{figure*}

With the large number of galaxies made accessible by the 
SDSS, it is possible to assess empirically how 
different galaxy properties,
such as stellar mass, stellar surface density, age, metallicity and
star formation rate map onto
the [O\,{\sc iii}]/H$\beta$ vs. [N\,{\sc ii}]/H$\alpha$ plane.
We have examined trends for a number of such quantities
(Charlot et al. 2008, in preparation)
and found the most relevant one for our purposes here
to be the equivalent width of the H$\alpha$ 
emission line.  It is not sufficient, however, simply to examine how
EW(H$\alpha$) varies with the diagnostic ratios in the
BPT diagram, in isolation from other physical parameters.
Rather, it is more instructive to consider EW(H$\alpha$)
\textit{differentially} relative to other `similar' galaxies,
where the meaning of `similar' depends on the issues
under investigation 
(see, for example, the discussion in Kauffmann et al. 2006).

In our case, the equivalent width of the H$\alpha$ emission line
is closely related to the specific star formation rate,
(SFR/$M_{\star}$, where $M_{\star}$ is the
assembled stellar mass), and both quantities then obviously
depend on $M_{\star}$ (Brinchmann et al. 2004). 
In addition, there is also a small residual trend 
between EW(H$\alpha$) and $d_A$ at fixed $M_{\star}$.
We removed the dependency on stellar mass and $d_A$ by calculating
for each galaxy the difference 
$\Delta{\rm EW(H}\alpha) = \log{\rm EW(H}\alpha) 
- \log \langle {\rm EW(H}\alpha) \rangle$ 
where $\langle {\rm EW(H}\alpha) \rangle$ is the median value
of the H$\alpha$ equivalent width in galaxies of the same 
mass and $d_A$.\footnote{In reality, the specific star formation rate
SFR/$M_{\star}$ depends on both $M_{\star}$ and the stellar
surface mass density $\mu_{\star}$ (Kauffmann et al. 2006).
However, once luminous bulges 
with $\mu_{\star} > 3 \times 10^8\,M_{\star}$~kpc$^{-2}$ 
are excluded, as we have done here, the main dependence is on $M_{\star}$.  }
Thus the quantity $\Delta$EW(H$\alpha$) measures the excess,
or deficit, of star formation activity in a galaxy relative
to the value typical of similar galaxies.\footnote{$\Delta$EW(H$\alpha$)
is also sensitive to differences in the degree of absorption of Lyman
continuum photons by dust \emph{within} an H\,{\sc ii} region,
but this is probably a second order effect.}
Mapping $\Delta$EW(H$\alpha$) onto the 
[O\,{\sc iii}]/H$\beta$ vs. [N\,{\sc ii}]/H$\alpha$ plane
by means of colour-coding, as in Figure~\ref{fig:bpt_variation},
shows very clearly that $\Delta$EW(H$\alpha$) is correlated
with $d_{\rm P}$, the offset from the ridge line
given by eq.\,(1).

Since our focus here is on empirical trends, 
we have chosen to concentrate on the equivalent
width of the H$\alpha$ line. 
But we have also verified that as long as average 
trends are taken out as above, 
the star formation rate (SFR), 
the specific star formation rate (SFR/$M_{\star}$), 
and the star formation rate per area (SFR/Area), 
all show trends with position in the BPT diagram 
which are similar to that of $\Delta$EW(H$\alpha$),
and with similar correlation strengths. 
We have also established that there
is no trend between $\Delta$EW(H$\alpha$),
defined as above, and the metallicity of
the galaxies in the range 
$7.2 <12 + \log \textrm{O/H} < 8.5$, 
where the oxygen abundance can be determined
directly from consideration of temperature
sensitive emission lines (the $T_{\rm e}$ method).

The pattern revealed by Figure~\ref{fig:bpt_variation} strongly
suggests that the high redshift galaxies lying well above the ridge
line in Figure~\ref{fig:bpt_with_highz} are experiencing unusually high levels
of star formation for their stellar mass compared to most local
galaxies. This interpretation is further strengthened by consideration
of Figure~\ref{fig:bpt_hiz_explanation1}, where we see that the
galaxies at $z > 1$ 
% studied by Erb et al. (2006a) and Kriek et al. (2007)
lie at, or beyond, the extremes of the locus of values
in the EW(H$\alpha$) vs. $M_{\star}$ plane occupied by SDSS galaxies.
[We could not include the measurements of DEEP2 galaxies by Shapley
et al. (2005) in the left panel of Figure~\ref{fig:bpt_hiz_explanation1} 
because those authors did not publish values of EW(H$\alpha$)].
It is quite remarkable that
the cut in EW(H$\alpha$) at a given mass, 
which involves \emph{no} line ratios, has such a close
correspondence on the position of the galaxies in 
the BPT diagram which \emph{only} uses line
ratios. 
We interpret this as evidence of a connection 
between star formation activity and 
the physical parameters determining the 
ratios of collisionally excited to recombination
lines in the H\,{\sc ii} regions.

\subsection{Interpretation}
\label{sec:interpretation}

In this section we explore further the connection 
between the specific star formation rate and the displacement
from the ridge line in the BPT diagram which we uncovered
in the preceding section.
Theoretical modelling of nebular emission from star-forming
galaxies (e.g. Charlot \& Longhetti 2001; Kewley et al. 2001a)
shows that a shift upwards in the BPT diagram can be achieved
by increasing either the ionisation parameter $U$ 
(which in the usual definition is
the ratio of the volume densities of ionising photons and particles)
or the dust-to-metals ratio, which measures the degree to which
heavy elements are depleted from the gas phase
and which Charlot \& Longhetti (2001)
denote with symbol $\xi_{\rm d}$ (or both).

\begin{figure}
  \centering
 % {\hspace*{0.05cm}
 % \includegraphics[angle=90,width=84mm]{model_trends_logU.ps}}
  {\hspace*{0.05cm}
  \includegraphics[angle=90,width=84mm]{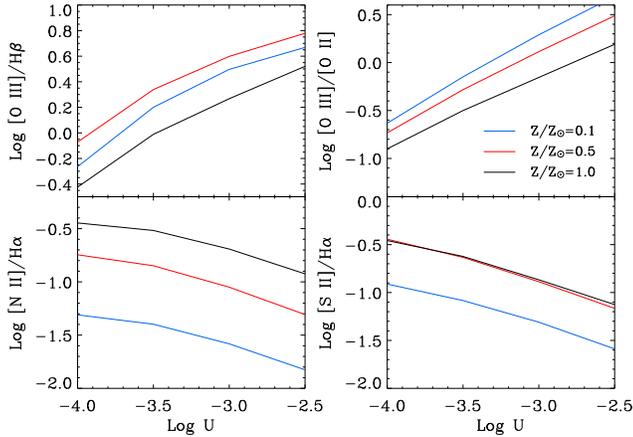}}
  \caption{Dependence of four diagnostic line ratios on the ionisation
    parameter $U$. Different colours correspond to different metallicities $Z$,
    as indicated. }
  \label{fig:line_ratios_U}
\end{figure}

\begin{figure}
  \centering
%  {\hspace*{0.05cm}
%  \includegraphics[angle=90,width=84mm]{model_trends_xsi.ps}}
  {\hspace*{0.05cm}
  \includegraphics[angle=90,width=84mm]{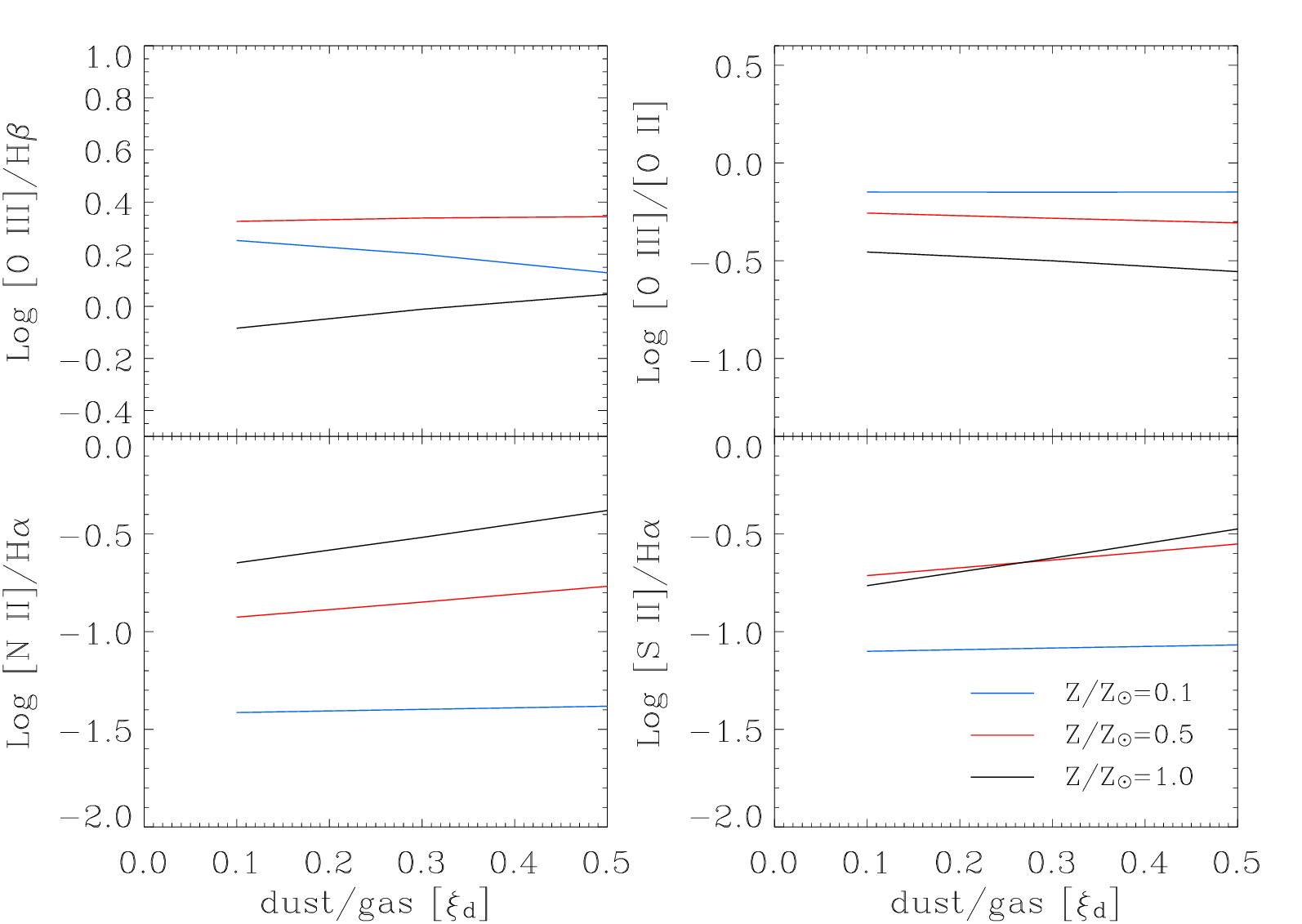}}
  \caption{Dependence of four diagnostic line ratios on the dust-to-metals
    ratio $\xi_{\rm d}$; the ionisation parameter was kept fixed at 
    $\log U = -3.5$. Different colours correspond to different metallicities $Z$,
    as indicated. } 
  \label{fig:models_vs_xsi}
\end{figure}

We have used the models of Charlot \& Longhetti (2001)
to illustrate the dependence of four emission line ratios
on the ionisation parameter $U$ in Figure~\ref{fig:line_ratios_U};
three representative values of metallicity $Z$ are considered.
As is well known, the [O\,{\sc iii}]/H$\beta$ ratio increases
with increasing $U$, while [N\,{\sc ii}]/H$\alpha$ decreases
(left-hand panels in  Figure~\ref{fig:line_ratios_U}); the strong
dependence of the latter on the metallicity partly helps to
resolve the metallicity-ionisation parameter degeneracy
in the BPT diagram. Also well understood (e.g. Penston et al.
1990) are the strong dependence of the [O\,{\sc iii}]/[O\,{\sc ii}]
ratio on $U$, with metallicity $Z$ being a second-order effect,
and the decrease of the ratio 
[S\,{\sc ii}]\,$\lambda\lambda 6716,6731$/H$\alpha$
with increasing $U$ (right-hand panels  
in  Figure~\ref{fig:line_ratios_U}).
The dependence of the same four emission line ratios
on the dust-to-metals ratio is, by comparison, much weaker
(Figure~\ref{fig:models_vs_xsi}); in general
the parameter $\xi_{\rm d}$ is less important than either $U$ or $Z$
in determining the values of the four 
emission line ratios considered.

\begin{figure*}
  \centering
  \includegraphics[width=147mm]{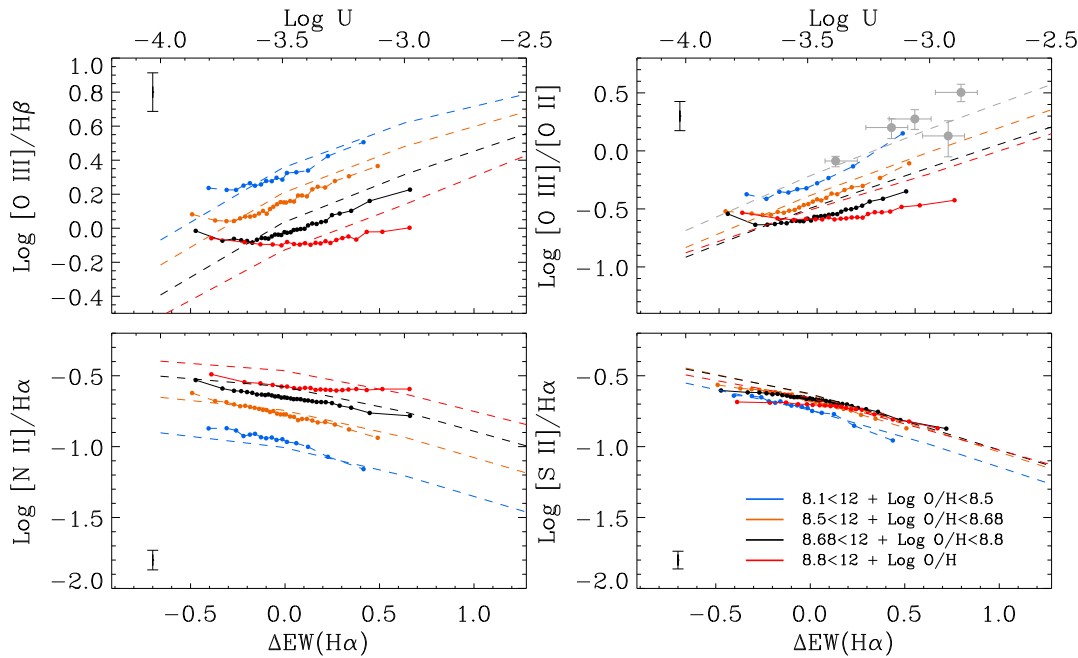}  
  \caption{Observed
    dependence of the four diagnostic line ratios on the 
    H$\alpha$ equivalent width offset $\Delta{\rm EW(H}\alpha)$
    in four different metallicity ranges, as indicated.
    For comparison,
    the dashed lines show the expected dependence of the emission line
    ratios on the ionisation parameter $U$ (given on the top $x$-axes),
    calculated using the Charlot \& Longhetti (2001) models. 
    The vertical bar in each panel indicates the $\pm 1 \sigma$
    spread of the 250 galaxies in each bin about the median value plotted.
    The spread is generally larger than the observational uncertainties
    and insensitive to the bin size adopted; we thus consider it to be
    a real spread in the emission line ratios at a given value of  
    $\Delta{\rm EW(H}\alpha)$. 
    In the top right-hand panel we have included (filled grey dots)
    the Lyman break galaxies in which Pettini et al. (2001) measured both
    [O \textsc{iii}]\,$\lambda 5007$ and  [O \textsc{ii}]\,$\lambda 3727$.
    The values of 
    [O \textsc{iii}]\,$\lambda 5007$/[O \textsc{ii}]\,$\lambda 3727$
    plotted here include a correction for dust
    appropriate to SDSS galaxies with similar metallicity (0.71), 
    and we estimated EW(H$\alpha$) from the values of EW(H$\beta$) 
    given by Pettini et al. (2001) using the average ratio 
    EW(H$\alpha$)/EW(H$\beta$) for star-forming galaxies in the SDSS (4.69).
        } 
  \label{fig:line_ratios_vs_dewha}
\end{figure*}

What is instructive is to consider the behaviour of the same
four line ratios as a function of the H$\alpha$ equivalent
width offset $\Delta{\rm EW(H}\alpha)$. To this end,
we have binned the SDSS DR4 galaxies (class SF) by metallicity
and in bins of  $\Delta{\rm EW(H}\alpha)$; each bin includes
250 galaxies. We have excluded galaxies with $d_{\rm A} < 1$
in order to avoid AGN contamination at the base of the AGN
plume in the BPT diagram. 
Figure~\ref{fig:line_ratios_vs_dewha} shows that the 
behaviour of the four line ratios in response to changes in
$\Delta{\rm EW(H}\alpha)$ is very similar to the response
to changes in $\log U$, pointing to a close connection between 
$\Delta{\rm EW(H}\alpha)$ and the ionisation parameter. 
If one were to take the [O\,{\sc iii}]/[O\,{\sc ii}]
ratio as a measure of $U$, as is often done, it would then
be difficult to escape the conclusion that $U$ and 
$\Delta{\rm EW(H}\alpha)$ are 
correlated.\footnote{Clearly,  it would be of interest 
in this context to also examine the 
[S\,{\sc ii}]$\lambda\lambda 6716, 6731$/[S\,{\sc iii}]$\lambda\lambda 9069, 9532$
ratio which Diaz et al. (1991) proposed as in 
independent estimator of the ionisation parameter.
Unfortunately, the SDSS spectra do not extend
sufficiently far into the infrared to measure the 
[S\,{\sc iii}] lines.}
The trends in Figure~\ref{fig:line_ratios_vs_dewha}  
also rule out a major contribution to the emission line ratios 
from shocks (Dopita \& Sutherland 1995). 
While we cannot make the same rigourous statement 
for the high redshift galaxies, their locations in the BPT diagram 
are not consistent with significant shock contributions 
(e.g.  see Figure~2 of van Dokkum et al. 2005).

We have already seen (section~\ref{sec:sfr_variation} 
and Figure~\ref{fig:bpt_variation}) that the quantity
$\Delta{\rm EW(H}\alpha)$ determines where
a galaxy falls in the BPT diagram
relative to the ridge line of `normal' star-forming galaxies.
It therefore seems at least plausible (although the argument
is indirect) that the offset of the high redshift star-forming 
galaxies in the BPT diagram is a reflection that their H\,{\sc ii}
regions are characterised by higher ionisation parameters
than those of local galaxies in the SDSS survey. 
Liu et al. (2008) reached a similar conclusion for their
smaller sample of more extreme SDSS galaxies.
A higher ionisation parameter is certainly consistent
with the high [O\,{\sc iii}]/[O\,{\sc ii}] ratios exhibited by
the $z\sim 3$ LBGs observed by Pettini et al. (2001), 
shown with grey dots in the top right panel of Figure~\ref{fig:line_ratios_vs_dewha}.
Adopting this inference as a working assumption,
we now proceed to consider which physical parameters 
may be responsible for the higher values of $U$.

\subsection{Possible reasons for a higher ionisation parameter}

We first consider the case where the H\,{\sc ii} regions are 
ionisation bounded (this assumption will be relaxed later).
The ionisation parameter at the edge of the 
effective Str{\"o}mgren sphere can then be written as
(e.g. Charlot \& Longhetti 2001):
\begin{equation}
  \label{eq:U_example}
     U(t) \propto \left [ Q(t) \, n_{\rm H} \,  \epsilon^2  \right ]^{1/3}
\end{equation}
where $Q(t)$ is the rate of hydrogen ionising photons (s$^{-1}$), 
$n_H$ is the number density of hydrogen (cm$^{-3}$), 
and $\epsilon$ is the volume-filling factor of the gas
(i.e. the ratio of the volume-averaged hydrogen density to $n_{\rm H}$).
Thus, an increase in $U$ can have one of three sources, which we now
consider in turn. Specifically, we have calculated the effects on
relevant emission line ratios of altering the parameters $Q(t)$, 
$n_{\rm H}$ and $\epsilon$; in order to do so, we have combined 
the output of our population synthesis models described in 
Section~\ref{sec:theor_models} with the photoionisation code
{\sc cloudy} (version 07.02---Ferland et al. 1998).

\subsubsection{Rate of ionising photons}
\label{sec:the_rate_of_Q}

The rate of ionising photons for a stellar population 
is a function of the composite stellar spectrum at ultraviolet 
wavelengths, which in turn depends on the mix of stellar
spectral types that contribute. Thus, while there may
be short-term fluctuations due to the evolutionary timescales
of stars of different masses, in the case of continuous 
star formation  the parameter $Q(t)$ quickly reaches
an equilibrium value.

\begin{figure}
  \centering
%  {\hspace{-0.35cm}\includegraphics[angle=90,width=86mm]{dp_time_evolution.ps}} 
  {\hspace{-0.35cm}\includegraphics[angle=90,width=86mm]{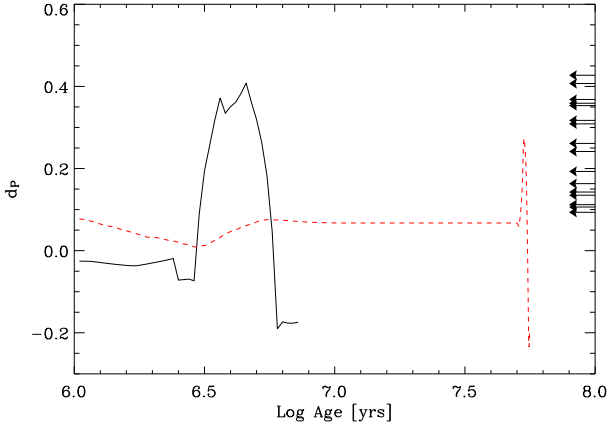}} 
  \caption{Time evolution of the parameter 
  $d_{\rm P}$ (which measures
     the offset perpendicular to the ridge line 
     in Figure~\ref{fig:bpt_with_highz})
   following an instantaneous burst of star formation (continuous line) 
   and a burst of constant star formation lasting 50\,Myr (dashed line).
   The models shown here are for solar metallicity.
   The arrows indicate the values of $d_{\rm P}$ appropriate to
   the $z > 1$ star-forming galaxies in Figure~\ref{fig:bpt_with_highz}.
  }
  \label{fig:o3_hb_burst}
\end{figure}

In our case, what is important is the ratio of photons that can ionise
O$^+$ to the hydrogen ionising photon; for an instantaneous burst,
this ratio decreases with time after the burst, except for a brief
increase during the Wolf-Rayet phase.  However, in the case of a more
protracted star formation episode, the ratio settles to a constant
value.  Figure~\ref{fig:o3_hb_burst} shows that even in the case of a
relatively short-lived burst of star formation, lasting only 50\,Myr,
temporary fluctuations in the value of the parameter $d_{\rm P}$
(which we recall measures the offset perpendicular to the ridge line
given by eq.~\ref{eq:bpt_ridge}) are washed out, so that $d_{\rm P}$
varies little with time until the end of the burst when it suddenly
drops. The majority of this change in $d_{\rm P}$ is due to a change
in the [O\,{\sc iii}]/H$\beta$ ratio. The fact that even within an
H\,{\sc ii} region stars are not all strictly coeval will further
smooth out the response of $d_{\rm P}$ to short-lived phases of
stellar evolution.

The situation is analogous to that already discussed in
Section~\ref{sec:bursts_effect} when we modelled the He\,{\sc
  ii}\,$\lambda 1640$ emission line. Our modelling shows that the
effects of bursts on $d_{\rm P}$ and [O\,{\sc iii}]/H$\beta$ are even
smaller than on EW($\lambda 1640$). Only when observing an H\,{\sc ii}
region at a very special time, such as 1\,Myr after the onset star
formation, do we expect to find elevated values of $d_{\rm P}$, but
this possibility is implausible when considering the integrated
spectrum of a whole galaxy.

Of course, an obvious way to increase the [O\,{\sc iii}]/H$\beta$
ratio is to appeal to a change in the initial mass function: if we
arbitrarily increased the relative numbers of the most massive stars,
a harder far-UV spectrum would be produced.  However, so far no direct
evidence has been found for such a change in the IMF in high redshift
star-forming galaxies. In the few cases where they have been recorded
at sufficient signal-to-noise ratio, the details of their UV spectra
are well reproduced by models with a Salpeter slope for stellar masses
in excess of $\sim 10\,M_{\odot}$ (e.g. Pettini et al. 2000; Steidel
et al. 2004; Erb et al. 2006a).

\subsubsection{Hydrogen density}
\label{sec:hydrogen_density}

Next we investigate the possibility that the higher ionisation
parameter in high redshift star-forming galaxies (our
working assumption) is due to higher densities of their
H\,{\sc ii} regions. We can address this point
using the well known dependence on the electron 
density, $n_{\rm e}$, of the ratio
of the S$^+$ emission lines,
[S\,{\sc ii}]\,$\lambda 6716$/[S\,{\sc ii}]\,$\lambda 6731$,
which varies from $\sim 1.5$ to $\sim 0.5$
as $n_{\rm e}$ increases from 1 to $10^4$\,cm$^{-3}$
(Osterbrock 1989). 

\begin{figure*}
  \vspace*{-0.25cm}
  \centering
%  {\hspace{-0.25cm}\includegraphics[width=184mm]{n_e_s2_vs_dp.ps}}
  {\hspace{-0.25cm}\includegraphics[width=184mm]{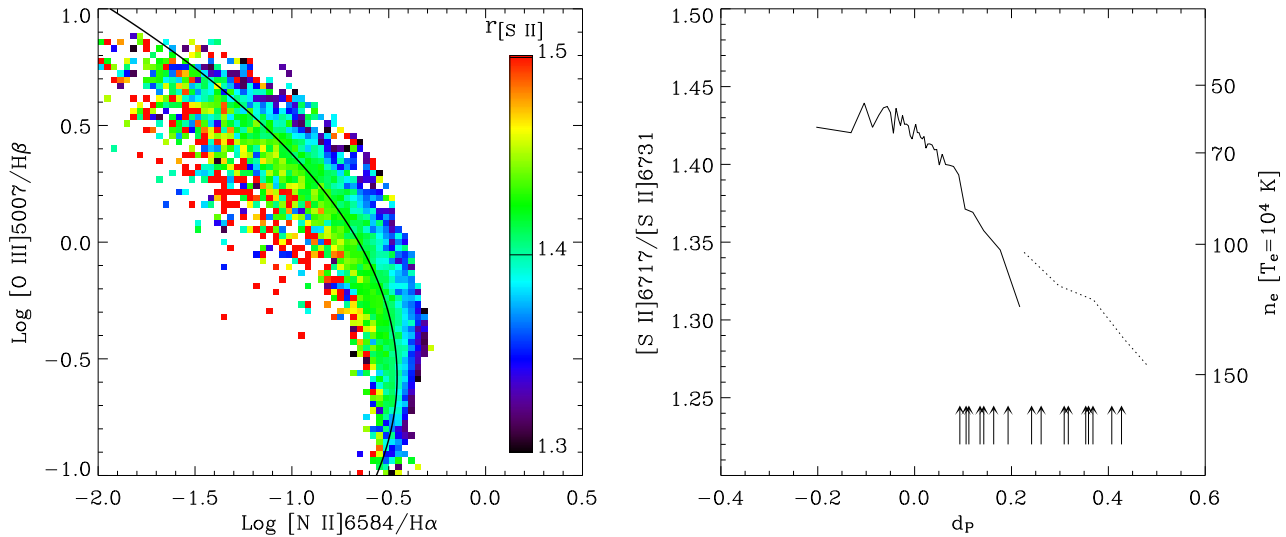}}
  \caption{
  \emph{Left: } When colour-coded according to the median
    value of the [S\,{\sc ii}]\,$\lambda 6716$/[S\,{\sc ii}]\,$\lambda 6731$
    ratio, galaxies separate on the BPT diagram, even though 
    the range of median values of this ratio is relatively small.~
   \emph{Right: } The density sensitive
     [S\,{\sc ii}]\,$\lambda 6716$/[S\,{\sc ii}]\,$\lambda 6731$
     ratio plotted vs. the parameter $d_{\rm P}$ which measures
     the offset perpendicular to the ridge line 
     in Figure~\ref{fig:bpt_with_highz}.
     The continuous line shows values 
     for SDSS DR4 galaxies of class SF, while the dotted line
     is for galaxies classified as `Composite'. 
     The values plotted are
     the medians in bins of 1000 galaxies each. 
     The right-hand $y$-axis shows the conversion to 
     the electron density $n_{\rm e}$
     for a fiducial temperature $T_{\rm e}=10^4$\,K.
     The arrows are plotted at the values of $d_{\rm P}$ 
     measured for the $z > 1$ star-forming galaxies in 
     Figure~\ref{fig:bpt_with_highz}. 
     The  [S\,{\sc ii}]\,$\lambda 6716$/[S\,{\sc ii}]\,$\lambda 6731$ ratio
      has not been measured with sufficient accuracy in any of
      these galaxies \emph{individually} 
      (hence the use of the arrow symbol), although
      from composite spectra values of $\sim 1.0 - 0.7$ 
      ($n_{\rm e} \sim 1000$\,cm$^{-3}$) seem typical.
       }
  \label{fig:sii_ratio_bpt}
\end{figure*}

In the left-hand panel of Figure~\ref{fig:sii_ratio_bpt} star-forming
galaxies from the SDSS DR4 have been colour coded according to their
[S\,{\sc ii}]\,$\lambda 6716$/[S\,{\sc ii}]\,$\lambda 6731$ ratio
which clearly differentiates them in the BPT diagram, even though the
ratio apparently varies by only a small fraction of the allowed range,
from [S\,{\sc ii}]\,$\lambda 6716$/[S\,{\sc ii}]\,$\lambda 6731 = 1.5$
to 1.3.  Within this restricted range, galaxies with smaller ratios
(higher $n_{\rm e}$) are preferentially found above the ridge line
defined by equation~(\ref{eq:bpt_ridge}).  It is therefore no surprise
that an inverse correlation between [S\,{\sc ii}]\,$\lambda
6716$/[S\,{\sc ii}]\,$\lambda 6731$ and the parameter $d_{\rm P}$ is
evident in the right-hand panel of Figure~\ref{fig:sii_ratio_bpt}.  We
constructed this plot by calculating the median [S\,{\sc
  ii}]\,$\lambda 6716$/[S\,{\sc ii}]\,$\lambda 6731$ ratio in bins of
$d_{\rm P}$ such that each bin included 1000 galaxies.  Again we have
excluded galaxies with $d_{\rm A} < 1$ in order to avoid AGN
contamination.  Although the trend between the median values is
clear-cut, we note that there is a substantial spread, of the order of
$\pm 0.1$\,dex (that is spanning the full range of the plot in the
Figure), in the values of [S\,{\sc ii}]\,$\lambda 6716$/[S\,{\sc
  ii}]\,$\lambda 6731$ measured in galaxies within each bin.  While at low
values of $d_{\rm P}$ this scatter is consistent with being due to
measurement error, at higher values it is likely that the ratio of
the two sulphur lines responds to other parameters apart from those
determining $d_{\rm P}$.

For an assumed electron temperature $T_{\rm e} = 10^4$\,K
(we do not have the means to measure $T_{\rm e}$ directly
but the temperature dependence is not severe), the median
values of 
[S\,{\sc ii}]\,$\lambda 6716$/[S\,{\sc ii}]\,$\lambda 6731$
plotted in Figure~\ref{fig:sii_ratio_bpt} imply a range
of electron density $n_{\rm e} \simeq 50 - 100$\,cm$^{-3}$.
Given the weak dependence of the ionisation parameter on density
(eq.~\ref{eq:U_example}), the corresponding range in $U$
amounts to only $\sim 0.06$\,dex.
However, the conditions at high redshift are more extreme.
Although there has yet to be a targeted study of the 
[S\,{\sc ii}] lines in high-$z$ star-forming galaxies, 
the indications from available data 
at $z \simeq 2$ (e.g. Shapley et al. 2004;
Erb et al. 2006a---see also Figure 3 of Pettini 2006) are that  
[S\,{\sc ii}]\,$\lambda 6716$/[S\,{\sc ii}]\,$\lambda 6731
\simeq  1.0 - 0.7$, corresponding to densities 
$n_{\rm e} \sim 10^3$\,cm$^{-3}$.
Such high densities are rarely encountered even in local starbursts
(e.g. Kewley et al. 2001b) and would account for a significant
increase---by about 0.5 dex in $U$---compared
to values typical of SDSS star-forming galaxies.
The indication from Figure~\ref{fig:sii_ratio_bpt} is
that such high densities may well be able to account
for the values of $d_{\rm P}$  exhibited by $z > 1$ star-forming
galaxies in Figure~\ref{fig:bpt_with_highz}.

\subsubsection{Volume filling factor}
\label{sec:volume_filling_factor}

While this parameter has potentially the most effect on the
ionisation parameter (see eq.~\ref{eq:U_example}), it is
also the one about which we know the least.
In H\,{\sc ii} regions of the Milky Way 
$\epsilon \sim 0.04$ (Kennicutt 1984), but we have 
little information on what determines this value 
and on how it may vary with other characteristics
of the H\,{\sc ii} regions. We have no grounds at present
for assessing whether it would greater (or smaller) 
in actively star-forming galaxies at high redshift.

\subsubsection{Density-bounded models}
\label{sec:density-bounded}

Equation~(\ref{eq:U_example}) assumes that the H\,{\sc ii} regions
are ionisation bounded; relaxing this assumption can have 
similar effects to an increase of the ionisation parameter.
Specifically, Binette, Wilson \& Storchi-Bergmann (1996) 
considered a composite model consisting of a mixture of 
radiation- and matter-bounded clouds and showed that altering
the proportion of the solid angle subtended by the two components,
(their parameter $A_{\rm M/I}$), closely mimics changes in
the ionisation parameter as far as the emission line ratios 
considered here are concerned. Similarly,
Giammanco, Beckman, \& Cedr{\'e}s (2005) 
calculated that an escape 
fraction of hydrogen ionising photons
$f_{\rm esc} = 50$\% from an H\,{\sc ii} region 
would be equivalent to an increase in $U$ 
by one order of magnitude compared to an analogous
H\,{\sc ii} region with zero escape fraction.

Could the positions of the high redshift galaxies in the BPT
diagram be a symptom that their H\,{\sc ii} regions
are `blistered' and allow a higher fraction of 
Lyman continuum photons to escape? 
This possibility is difficult to assess quantitatively
without more detailed modelling. From the work by
Binette et al. (1996) and Giammanco et al. (2005)
we estimate that considerable escape fractions, 
$f_{\rm esc} > 10$\%, would be required to explain
the values of $d_{\rm P}$ exhibited by the galaxies at
$z > 1$ in Figure~\ref{fig:bpt_with_highz}.
Observationally, the direct measurement of 
$f_{\rm esc}$ has proved to be very challenging. 
So far, attempts to detect Lyman continuum emission
in nearby galaxies
have produced null results in all cases except possibly one
(see Bergvall et al. 2006 and references therein; Grimes et al. 2007).
The best limits at $z \simeq 1$ imply an upper
limit $f_{\rm esc} < 0.15$ (Siana et al. 2007),
consistent with the detection of near 100\% escape
fraction in two out of 14 Lyman break galaxies 
at $z \sim 3$ and negligible values in the other
12 by Shapley et al. (2006). The difficulty in 
interpreting all of these measurements is that 
we remain essentially ignorant of what factors
affect $f_{\rm esc}$; until these are understood,
it is highly speculative to draw conclusions on the 
average escape fraction of Lyman continuum photons
from galaxies on the basis of a few detections.
Nevertheless, there have been claims that
$f_{\rm esc}$ does evolve in the sense that it was
higher at earlier times (Inoue, Iwata, \& Deharveng
2006---see also Bolton et al. 2005 for less direct 
arguments), and therefore this could also be an important
factor contributing to the offset of the high-$z$
galaxies in the BPT diagram.

\subsection{Conclusions and implications}
\label{sec:implications}

Recapitulating the main points of this section,
we have found that SDSS galaxies with elevated levels
of specific star formation (SFR/$M_{\star}$) 
tend to be offset in the BPT diagram 
towards higher values of the collisionally excited
lines relative to the recombination lines.
This is the same sense of the larger offset 
apparently exhibited by galaxies at $z > 1$, 
although the numbers of such galaxies where all
four nebular lines,
[O\,{\sc iii}], [N\,{\sc ii}], H$\alpha$ and 
H$\beta$,  have been measured so far is still
very small.
The response of the diagnostic emission line ratios
to the excess, or deficit, of H$\alpha$ equivalent width
(compared to galaxies of similar stellar mass) mirrors
their response to changes in the ionisation parameter,
suggesting that H\,{\sc ii} regions of high redshift
star-forming galaxies may be characterised by systematically
higher values of the ionisation parameter $U$ than galaxies
at the present time. Among the physical properties that
may lead to higher values of $U$,  we have identified high
electron densities, $n_{\rm e} \sim 1000$\,cm$^{-3}$, and 
a non-zero escape fraction of Lyman continuum photons
(from nebulae which are at least partially density-bounded)
as the most likely contributors on the basis of the still very limited
body of available data. As the number of high-$z$ galaxies with
accurate measures of the optical emission line ratios increases,
it will become possible to test these ideas further with more
detailed photoionisation models than warranted at present.

\begin{figure*}
  \centering
  \includegraphics[width=174mm]{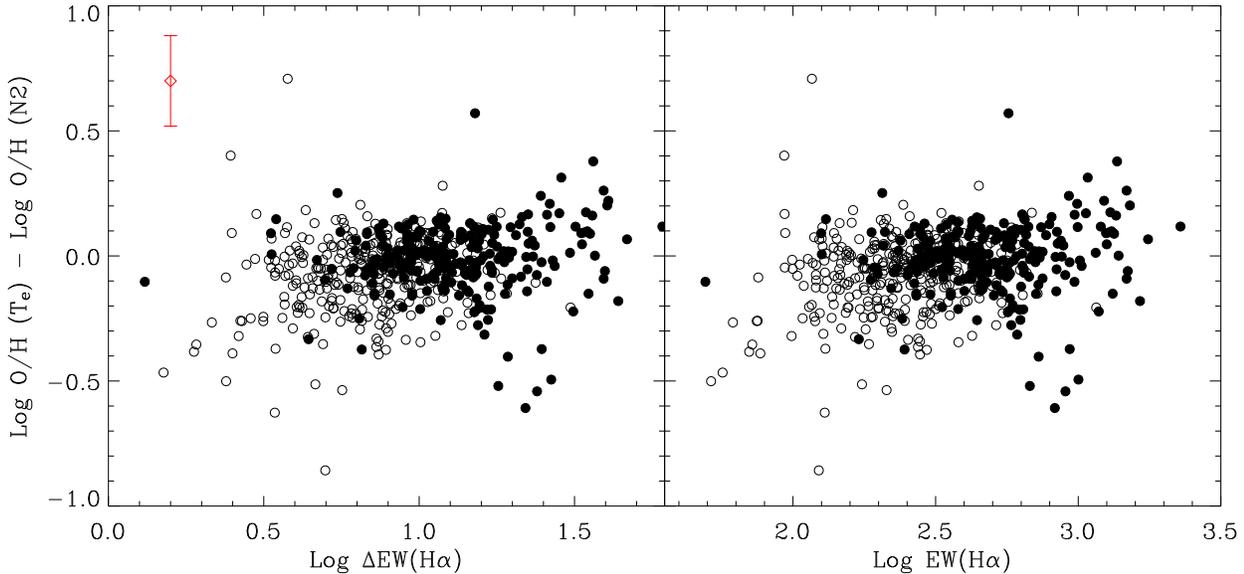}    
  \caption{\textit{Left:} Difference
    between O/H estimates from the $T_{\rm e}$ and $N2$ methods
    as a function of the differential H$\alpha$ equivalent width,
    $\Delta {\rm H}\alpha$, for 584 star-forming galaxies from the SDSS
    DR4. 
    Filled dots indicate galaxies where the temperature-sensitive
    [O\,{\sc iii}]\,$\lambda 4363$ auroral line is detected with
    $S/N > 10$, while the open circles are for galaxies where
    the line is measured with an accuracy of between 10 and 20\%.
    The error bar in the top left-hand corner indicates the
    median uncertainty in 
    $\log$\,[O/H\,($T_{\rm e})$/O/H\,($N2$)],
    calculated by combining in quadrature the error in
    O/H\,($T_{\rm e})$ (mostly due to the accuracy in the 
    measurement of the weak [O\,{\sc iii}]\,$\lambda 4363$
    emission line) with the $1 \sigma$ scatter of 0.18\,dex
    in the $N2$ calibration by Pettini \& Pagel (2004).~
    \textit{Right:} The oxygen abundance difference plotted
    as a function of the equivalent width of H$\alpha$.
    The only SDSS galaxies where [O\,{\sc iii}]\,$\lambda 4363$
    can be measured reliably are those with
    EW(H$\alpha) \simgt 100$\,\AA.
    }
  \label{fig:dewha_deltaOH}
\end{figure*}

One of the motivations for clarifying the origin of the 
offset of high-$z$ galaxies in the BPT diagram was the 
concern that such an offset may bias metallicity
measures based on the ratios of strong emission lines
(the so-called strong-line methods).
Among the many approaches to determining the
metal content of high redshift galaxies, 
the calibration of the $N2$ index 
[$N2 \equiv \log$\,([N\,{\sc ii}]\,$\lambda 6583/{\rm H}\alpha)$]
as a function of the oxygen abundance by Pettini \& Pagel (2004)
has turned out to be the most convenient to apply to high redshift
star-forming galaxies for practical reasons (see Pettini 2006
for a comprehensive discussion).
As we saw in Figure~\ref{fig:line_ratios_U},
the [N\,{\sc ii}]/H$\alpha$ ratio, exhibits a monotonic
decrease with increasing ionisation parameter at a 
fixed metallicity. If high redshift galaxies have systematically
higher values of $U$, as argued here, what effect will that
have on the calibration of O/H vs. the $N2$ index?

As a first step towards addressing this question,
we compare the oxygen abundance
deduced from the $N2$ index with that from the
more rigorous effective temperature 
(or $T_{\rm e}$) method and examine whether 
the difference between these two values
of O/H is correlated with the H$\alpha$ equivalent 
width offset $\Delta {\rm EW(H}\alpha$) as defined 
in Section~\ref{sec:sfr_variation}. 
The $T_{\rm e}$ method could be applied to only 584
galaxies from SDSS DR4 (of class SF) since it relies
on the detection of the weak [O\,{\sc iii}]\,$\lambda 4363$
emission line. The measurement of this line, and therefore
the determination of the nebular temperature, can be affected
considerably from the precise placement of the underlying
stellar continuum; for this reason we limited ourselves to
considering galaxies where [O\,{\sc iii}]\,$\lambda 4363$ is
detected with a signal-to-noise ratio $S/N > 5$.
In deducing O/H we used the fitting functions of
Izotov et al. (2006) for the $T_{\rm e}$ method and
the calibration of the $N2$ index by Pettini \& Pagel (2004).

As can be seen from Figure~\ref{fig:dewha_deltaOH},
the two estimators of O/H are closely related, which 
comes as no surprise since Pettini \& Pagel calibrated
the $N2$ index on $T_{\rm e}$ measurements of the
oxygen abundance in nearby galaxies. What is interesting
for our present enquiry is the fact that, unlike other
parameters we have considered in this work,
there seems to be no dependence 
on $\Delta {\rm EW(H}\alpha)$
of the difference in
O/H between the two methods, 
at least for the galaxies
where [O\,{\sc iii}]\,$\lambda 4363$ is detected with
$S/N > 10$ (filled dots in Figure~\ref{fig:dewha_deltaOH}).
If there are any trends with $\Delta {\rm EW(H}\alpha)$,
they are masked by the larger uncertainties in the determination
of the oxygen abundance with either method (indicated by the
error bar in the top left-hand corner of Figure~\ref{fig:dewha_deltaOH}). 

The lack of a clear trend in Figure~\ref{fig:dewha_deltaOH} 
between the difference in the two measures of the oxygen abundance 
on the one hand, and $\Delta$EW(H$\alpha$) on the other, may be
surprising at first sight. If the $N2$ index varies with $U$ at a
fixed metallicity (as we saw in the lower left-hand plot in 
Figure~\ref{fig:line_ratios_U}) and if $\Delta$EW(H$\alpha$)
is related to $U$, as we have argued, then might we not expect
to see larger differences in 
$\log$\,O/H($T_{\rm e}) - \log$\,O/H($N2$) at higher
values of $\Delta$EW(H$\alpha$)?
The answer to this apparent puzzle lies in the fact that the
$N2$ index of Pettini \& Pagel (2004) is \emph{empirically} calibrated;
a range of galaxies with different values of the ionisation
parameter were used in the calibration, so that a given value of $N2$ 
does not correspond to a single value
of $U$.  The shallow slope of the O/H vs. $N2$ relation
then results in relatively minor offsets in O/H($N2$) between
galaxies with systematically different values of $U$,
as already discussed by Pettini \& Pagel (2004).
Such offsets are apparently washed out in Figure~\ref{fig:dewha_deltaOH}
by other sources of error or intrinsic dispersion.

While the agreement between the two oxygen abundance
indicators in Figure~\ref{fig:dewha_deltaOH} is 
reassuring\footnote{We have also verified that the  same conclusion 
holds for the O3N2 estimator of Pettini \& Pagel (2004).}
(although not unexpected, as explained), 
it must be borne in mind that we have been able to perform this
test for only a small subset of SDSS galaxies. The galaxies
where the [O\,{\sc iii}]\,$\lambda 4363$ auroral line
can be detected with
high significance in SDSS spectra are overwhelmingly low mass,
low metallicity, high $T_{\rm e}$ systems 
with high values of EW(H$\alpha$),
as can be readily appreciated from inspection
of the right-hand panel of Figure~\ref{fig:dewha_deltaOH}.
Whether the agreement extends to the higher metallicities
more typical of high-$z$ star-forming galaxies is a significantly
more difficult question to address (e.g. Bresolin 2006, 2007 and 
references therein).

\section{Overall Summary}
\label{sec:conclusions}

In this paper we have brought together the large body of
observations of local galaxies from the SDSS DR4, and
state-of-the-art population synthesis models---coupled to 
the photoionisation code {\sc cloudy}
and including recent reassessments of the 
contributions from Wolf-Rayet stars---in order to interpret
some key spectral features seen in the spectra of high
redshift star-forming galaxies. 
Our most important findings are as follows.

1. A detectable stellar He\,{\sc ii}\,$\lambda 1640$ emission
line, with an equivalent width of $0.5-1.5$\,\AA, is expected
to be present in the spectra of galaxies undergoing 
continuous star formation. The equivalent width of the 
line is stable to fluctuations in the star-formation rate,
because, to a first approximation,
the luminosities of the emission line and the 
underlying continuum are both
determined by the number of massive stars.

2. In our models, the most important parameter 
driving EW($\lambda 1640$) is the metallicity $Z$,
which affects both the fraction of W-R to O-type stars
and the line luminosity in a W-R star of a given (sub-)type:
at lower metallicities, fewer massive stars enter the W-R stage
and the line is intrinsically weaker. 
Empirical  data of relevance to the calibration of
$L(\lambda 1640)$ vs. metallicity are still sparse, as
observations of individual W-R stars have so far been
largely limited to galaxies in the Local Group.
However, we find an encouraging agreement between the
value of EW($\lambda 1640$) calculated with our models
using current estimates of the typical 
metallicity $Z \simeq 0.5 Z_{\odot}$
of bright (${\cal R} < 25.5$) UV-selected galaxies 
at redshifts $z = 2- 3$, 
and the value measured from the composite
spectrum of hundreds of Lyman break galaxies.
This result opens up the possibility of using in future
the He\,{\sc ii}\,$\lambda 1640$ emission line 
to probe the wind properties of massive stars 
in high redshift galaxies.

3. We have uncovered a relationship between the 
specific star formation rate and the position of an SDSS
galaxy in the diagnostic [O\,{\sc iii}]/H$\beta$ vs.
[N\,{\sc ii}]/H$\alpha$ nebular emission line diagram
(the BPT diagram). Galaxies with excess H$\alpha$
equivalent width compared to the median for their
stellar mass are displaced towards larger values
of the two emission line ratios in the BPT diagram
(and \textit{vice versa}). 

4. The responses of several emission line ratios
to this parameter, which we denote as $\Delta {\rm EW(H}\alpha$),
is qualitatively similar to their responses to changes in the 
ionisation parameter.
This observation suggests that the offsets in the BPT
diagram exhibited by the few star-forming galaxies at
$z > 1$ in which the same line ratios have been 
measured, may be an indication that their H\,{\sc ii}
regions are characterised by systematically larger
values of the ionisation parameter $U$ than most
local star-forming galaxies. 
Liu et al. (2008) have recently reached a similar
conclusion from their analysis of a somewhat 
different sample of SDSS galaxies.

5. If this is the case, we identify higher electron densities
and larger escape fractions of hydrogen ionising photons
as two physical parameters which may be responsible for
the elevated values of $U$ and, ultimately, the offsets
in the BPT diagram. 

6. To first order, it appears that the systematically higher
values of $U$ may have only a relatively minor effect on the
determination of the oxygen abundance from strong-line
methods, possibly because of the larger inherent scatter
in the calibrations of the most commonly used strong-line indices. 

Finally, we look forward to the forthcoming availability
on large telescopes of near-infrared
spectrographs offering multi-object capabilities
and/or wide spectral coverage; such new instrumentation
will allow significant 
progress to be made in the investigation of the 
questions we have begun to explore here.

\section{Acknowledgements}

We gratefully acknowledge valuable discussion 
with Paul Crowther, Dawn
Erb, Alice Shapley and Thierry Contini. 
JB acknowledges the receipt of 
FCT grant SFRH/BPD/14398/2003. 
An anonymous referee made
a number of valuable suggestions 
which improved the paper.

Funding for the SDSS and SDSS-II has been provided by the Alfred P.
Sloan Foundation, the Participating Institutions, the National Science
Foundation, the U.S. Department of Energy, the National Aeronautics
and Space Administration, the Japanese Monbukagakusho, the Max Planck
Society, and the Higher Education Funding Council for England. The
SDSS Web Site is \texttt{http://www.sdss.org/}.

The SDSS is managed by the Astrophysical Research Consortium for the
Participating Institutions. The Participating Institutions are the
American Museum of Natural History, Astrophysical Institute Potsdam,
University of Basel, University of Cambridge, Case Western Reserve
University, University of Chicago, Drexel University, Fermilab, the
Institute for Advanced Study, the Japan Participation Group, Johns
Hopkins University, the Joint Institute for Nuclear Astrophysics, the
Kavli Institute for Particle Astrophysics and Cosmology, the Korean
Scientist Group, the Chinese Academy of Sciences (LAMOST), Los Alamos
National Laboratory, the Max-Planck-Institute for Astronomy (MPIA),
the Max-Planck-Institute for Astrophysics (MPA), New Mexico State
University, Ohio State University, University of Pittsburgh,
University of Portsmouth, Princeton University, the United States
Naval Observatory, and the University of Washington.

\appendix

\section{The luminosity weighted metallicity of the composite LBG spectrum}
\label{sec:lbg_metallicity}

\begin{figure}
  \centering
  \includegraphics[angle=90,width=84mm]{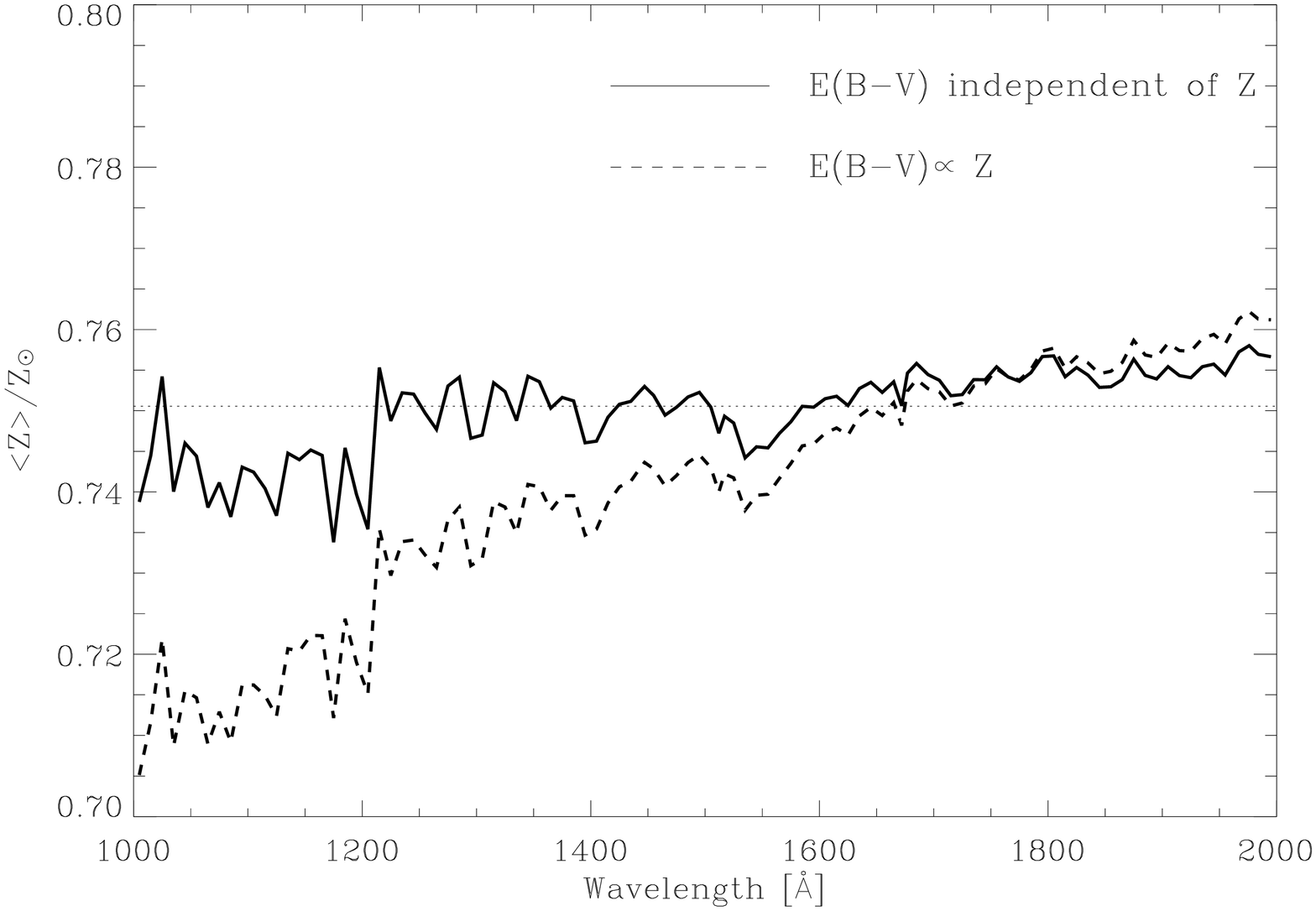}
  \caption{Luminosity-weighted metallicity of a composite spectrum
    of model Lyman break galaxies where the individual have
    been co-added after normalising to a common mode of the 
    signal between 1250\,\AA\ and 1500\,\AA, as in Shapley
    et al. (2003)---see text for further details. 
    The models spectra combined to produce the composite have an 
    average metallicity $\langle Z \rangle = 0.75\, Z_{\odot}$, 
    indicated by the dotted line.} 
  \label{fig:avg_lbg_zmet}
\end{figure}

As discussed in Section~\ref{sec:model_HeII}, 
current estimates of the nebular oxygen abundance in 
Lyman break galaxies are 
(O/H)$_{\rm LBG} \approx 0.5\, {\rm (O/H)}_{\odot}$
(e.g. Pettini et al. 2001; Erb et al. 2006a).
If this is taken to be representative of the population
of LBGs brighter than $R = 25.5$ (i.e. brighter than
$\sim 0.5\, L^{\star}$; Reddy et al. 2007) 
at $z \simeq 3$,  we find that there is reasonably good agreement
between the value of the equivalent width
of the He\,{\sc ii}\,$\lambda 1640$ emission line
calculated by our models and that measured from the composite
spectrum of 811 LBGs constructed by Shapley et al. (2003).

In the main text of the paper we cautioned that this agreement
may be somewhat fortuitous considering the significant spread in the
luminosity of the He\,{\sc ii} lines measured in stars classified
as belonging to the same Wolf-Rayet spectral sub-class. Given
the paucity of relevant data, the
spread is still poorly determined nor are its underlying causes
understood. Here we briefly consider another, more subtle,
source of possible confusion in comparisons with co-added
spectra of many galaxies; it is worthwhile pointing out such
potential complications because the signal-to-noise ratio of
single spectra is often too low for
individual analysis.

In constructing their composite of 811 LBGs, Shapley et al. (2003)
renormalised each spectrum to the common mode of the continuum 
flux between 1250\,\AA\ and 1500\,\AA. 
In principle, if there is significant range of continuum slopes
among the 811 galaxies and if the slope is related to the 
metallicity (which seems plausible), the procedure of normalising
to the mode may lead to a systematic bias in the signal
near the He\,{\sc ii}\,$\lambda 1640$ emission line
(because galaxies of a given metallicity may contribute
more than the average at a given wavelength).

We tested the magnitude of such a potential bias using a 
library of model spectra 
generated as discussed in Section~\ref{sec:theor_models}.
The library contains spectra produced by $\sim 10^5$ stochastic realisations  
of the population synthesis code by  Bruzual \& Charlot (2003).
The details of this grid of models have been described 
by Gallazzi et al. (2005) and Salim et al. (2005, 2007); 
we refer the interested reader to those papers for a full discussion 
of how they are generated. 
Briefly, the star formation history is characterised 
by two components: an underlying continuous star 
formation with an exponentially declining rate  
determined by the parameter 
$\tau$ (${\rm SFR} \propto \exp [-t_{\rm sf}/\tau]$), 
on which random bursts can be  superimposed. 
Dust obscuration is included so that the distribution 
of dust attenuation of the stellar birth clouds 
reproduces the distribution of H$\alpha$/H$\beta$ 
ratios of the star-forming galaxies in the SDSS, 
although for the present purposes  
the dust content is not very important.

From this library we select all galaxies which satisfy the LBG colour selection
criteria of Steidel et al. (2003). The models span a range of metallicity
$0.3 <Z/Z_{\odot} < 1.2$, with an average 
$\langle Z \rangle = 3/4 \,Z_{\odot}$.
The model spectra were added together using the same normalisation
as Shapley et al. (2003) and the luminosity-weighted metallicity was
calculated at each wavelength.  The results
of this exercise are illustrated in 
Figure~\ref{fig:avg_lbg_zmet}.
Two cases are considered: in one case the reddening is
independent of metallicity, whereas in the other $E(B-V)$
increases linearly with $Z$. The two simulated spectra are
normalised so that they match at 1640\,\AA. 
Inspection of Figure~\ref{fig:avg_lbg_zmet} shows that,
while there
is indeed a subtle bias, in either case the effect is so small 
as to be
inconsequential for most current applications.

\label{lastpage}

\end{document}